\documentclass{article}

\begin{document}

\title{Kinetic effects  of multi-component nucleation}
\author{Victor Kurasov}

\date{St.Petersburg State University,
\\
Department of Computational
Physics
\\e-mail : \\
Victor.Kurasov@pobox.spbu.ru}

\maketitle

\begin{abstract}

The process of multicomponent condensation
 is considered. The
theory taking into account several channels of nucleation is
constructed.   The analytical approximate description of the whole
condensation process is given. The specific phenomena of the
secondary nucleation is described analytically. The possibility of
the partial reverse condensation is discussed.

\end{abstract}

PACS:  05.20.Dd; 64.70.Fx; 64.60.Qb; 82.70.R

Keywords: Multicomponent condensation; Kinetics; Nucleation stage;
Vapor depletion;

\section{Introduction}

Multicomponent nucleation is widely spread in nature and the task
to
describe it theoretically is rather actual.
Creation of the classical theory of nucleation \cite{class} allows
to investigate the case of multicomponent nucleation. The
stationary rate of
multicomponent  nucleation was found in \cite{stauffer},
 solution of a stationary kinetic equation with
appropriate boundary conditions was given in \cite{tech}. These
results gave the opportunity to describe kinetics of the global
evolution during the whole phase transition. The final period of
evolution is known - it is the period of
the recondensation (coalescence) and it
is described by the Lifshic-Slezov theory \cite{Lifshic-Slezov}. But
coalescence is mainly the redistribution of surplus
substance\footnote{Certainly,
during the coalescence a small part of the surplus substance
is gradually going from the mother phase to the new phase.} and the
real transition of  practically all surplus substance occurs
earlier.
Before the coalescence, in fact, a  phase transition takes place.
To describe the phase transition it is necessary to determine
the total number of droplets appeared in the process of
nucleation.  This quantity is the quasiintegral of evolution
after the end of nucleation and before the coalescence.
 The characteristic times of coalescence ordinary strongly exceeds
the characteristic  times
of experimental observations and the
characteristic time of a practically all  surplus
substance consumption exceeds the time of nucleation
(i.e. the duration of the intensive appearance of droplets).
Then the problem to determine the number of droplets is the central
one and it will
be solved here when many channels of nucleation a important.

The problem to determine the global evolution until coalescence
for multicomponent case was investigated in \cite{stauffer},
\cite{preprint0}, \cite{preprint1}, \cite{preprint2},
\cite{physicaa94}. Recent publications \cite{djikaiev1},
\cite{djikaiev2} repeat\footnote{ But contain some wrong
statements, which will be corrected  here.} the  consideration made
in \cite{preprint1}, \cite{preprint2}, \cite{physicaa94}. The
mentioned publications gave the consideration of the most spread
case of multicomponent condensation when the droplets of the sole
composition will appear in the system. Then as it is noted in
\cite{preprint1}, \cite{preprint2}, \cite{physicaa94},
\cite{preprint0} the method of description of the nucleation can
be reduced by renormalizations to the one component
case. Therefore it was  sufficient to give in \cite{preprint1},
\cite{preprint2}, \cite{physicaa94}, \cite{preprint0} the
corresponding references to the one component  case \cite{4}.
Nevertheless in \cite{djikaiev1}, \cite{djikaiev2} in consideration
of the
binary condensation the method of solution of one component
situation was reproduced and unfortunately  it was done with many
errors. This requires to give here detailed analysis of this
problem.

Besides the consideration of nucleation with a sole composition of
components in droplets the case of multiple compositions  will be
considered here. To see that the consideration based only on the
sole composition isn't complete one we can imagine the following
simple example. Let us suppose that the system contains two
 immiscible substances and both substances in a vapor
phase are supersaturated. Then it is evident that the two separate
processes of condensation will take place\footnote{Certainly,
these processes can interact by a heat release effects, but when
these effects are small (it can be ensured by sufficient quantity
of passive gas) one will observe two absolutely separate processes
of the one component nucleation.}. As the result there will be
droplets of the first substance and of the second
substance\footnote{Due to the entropy of mixing there can be some
species of the other substance in  droplets of the given
dominating substance.}. So, it is evident that consideration of
multicomponent nucleation should contain the consideration of
multiple channels of droplets formation. It will be done below.
The effect of multiplicity of channels of nucleation is well
known, it is deeply investigated by  Ray \cite{Ray}. It is rather
strange that in \cite{djikaiev2} (where \cite{Ray} is considered as
the base for constructions) these conclusions of Ray are ignored.

From the first pont of view it
 seems that this problem can be solved by the simple
generalization of the case with the sole composition of substances
in droplets. This illusion is wrong. The situation when there is
absolutely no interaction between separate processes of nucleation
of droplets with different compositions is exclusive. Normally
there is such interaction and it has to be taken into account.
This interaction forms the main content of  multicomponent
nucleation.

To speak about kinetic effects of multicomponent nucleation one
has to specify external conditions. We shall consider the process
of the metastable phase decay. In the initial moment of time
 we have a metastable mixture
of gases. The stable phase is the liquid droplets phase, which will
gradually appear in the system. Liquid phase droplets will consume
metastable vapor molecules and the supersaturation of the mother
phase is gradually falling. It leads to the cessation of
appearance of new droplets and stops the process of nucleation.
This is the qualitative picture of metastable phase decay, which
will be described quantitatively.

We consider the free molecular regime of vapor consumption. In
\cite{PhysicaA95} it is shown how renormalizations can reduce the
situation with an arbitrary regime of vapor consumption to the
case of the free molecular regime. Below some corresponding remarks
will be given.

The absence of thermal effects of condensation is supposed.
Formally speaking this absence can be
attained  by sufficient quantity of a
passive gas. The  generalization
on the thermal effects of nucleation is rather evident and can be done
in a manner quite similar to \cite{Novos}.

We take the classical thermodynamic expression for the free energy
of a critical
embryo\footnote{Or some modified approach. As it can be seen below
only some general features of the free energy (such as smoothness
on the power of metastability of a mother phase)
are required.}. This requires that the
critical embryo has
to contain many molecules of every component. Only then we can use
the approximation of a liquid solution in the embryo. Otherwise
one can consider the molecules of a rare component as
heterogeneous centers, but the requirement to have many molecules
of all substances
in the critical embryo remains\footnote{
The theory for heterogeneous decay is given in \cite{book}.
}. Certainly, one can argue about the
discrepancy between theoretical and experimental results in
nucleation rates even in the one component homogeneous  case, but
it seems rather evident that this discrepancy can be eliminated by
some fitting (normalizing) constants, which are very smooth
functions of all ordinary thermodynamic parameters of the systems
including supersaturation and temperature. So, we can use the
classical expression for the stationary nucleation rate as a brick
for constructions made in this paper. It can be done
 because the variations in
thermodynamic parameters essential for nucleation will be very
small (details see later).

The system is supposed to be homogeneous in space. The unit volume
is considered, all energy like values are measured in units of a
mean thermal energy.

\section{Channels of nucleation}

\subsection{Asymptotic thermodynamic channels}

It is supposed that in the system there are $l$ different
substances, which can be found later in droplets of liquid. They
will be called as components. Different components are marked by
lower index. The molecular number densities are marked by $n_i$,
in the initial moment of time $t$ they are known and equal to
$n_i(0)$. The number of molecules of $i$-th component in the
embryo is marked by $\nu_i$, the total number of molecules  - by
$$
\vartheta
 = \sum_i
\nu_i\ . $$
 The embryo can be described by the set $\{\nu_i\}
\equiv \vec{\nu}$.  Then the concentration $\xi_i$ of component
inside the embryo will be\footnote{We  don't take into account
here the surface enrichment effects.  It will be done later. }
$$
\xi_i = \nu_i / \vartheta \ \ .
$$
 It can be done for every
component and the vector $\vec{\xi} = \{ \xi_i \} $ can be
introduced\footnote{Here one concentration is
absent due to $\sum_i \xi_i = 1$.}.
Then one can speak about the molecular number density
of a vapor of $i$-th component saturated over a liquid of
concentration $\vec{\xi}$ with a plane surface. It is marked by
$n_{ie}$. According to standard definitions
$$
n_{ie} =
n_{i\infty} \xi_i f_i(\vec{\xi}) \ \ ,
$$
 where $n_{i\infty}
\equiv n_{ie}(\xi_i = 1)$ and $f_i$ are activity coefficients.
Ordinary they satisfy conditions\footnote{We shall use in this paper two
sets of variables for the embryo description $\{\xi_i\} ,
\vartheta$ and $\{\nu_i\}$. Sometimes $\vartheta$ will be changed
by the volume of the embryo. So, it is clear what variables will
be fixed in partial derivatives.}
$$
f_i (\xi_i \rightarrow  1)
\rightarrow 1 ,  \ \ \partial f_i / \partial \xi_i (\xi_i \rightarrow
1 ) \rightarrow 0.
$$
 In approximation of ideal solution $f_i \equiv
1$.

The free energy of an embryo is marked by $F$. One can speak about
$F$ only for an embryo near the  state of initial equilibrium. For
the bulk liquid phase the free energy is given by $F_b =  - \sum_i
\nu_i \Delta \mu_i$ where the differences of chemical potentials
are given by $\Delta \mu_i = \ln(n_i/n_{ie})$, which implies that
the vapor phase can be regarded as an ideal
gas\footnote{Certainly, virial corrections can be introduced here
and this allows to describe many real situations.
}. Having extracted extensive variable $\vartheta$ one can rewrite
$F_b$ as
$$F_b = -  \vartheta \sum_i \xi_i \ln(n_i/n_{ie})$$
and extract the function
$$
B_b = - \sum_i \xi_i \ln(n_i/n_{ie}).
$$
On the base of $B_b$ one can formulate the necessary condition for
the possibility of nucleation - function $B_b$ has to be negative
at least at some $\xi$.

The form $F_b = \vartheta B_b (\xi)$ leads to existence of the
thermodynamic asymptotic channels of nucleation. These channels
are straight lines in plane $\vec{\xi},\vartheta$, they are
perpendicular to $\vec{\xi}$-plane. The coordinates $\xi_{i\ c}$
are extracted by conditions
$$
\frac{\partial B_b}{\partial \xi_i}
= 0, \ \ \ S_q  (B_b) \equiv \sum_{i,j} \frac{\partial B_b}{\partial
\xi_i \partial \xi_j } (\xi_i - \xi_{i\ c})(\xi_j - \xi_{j\ c})
\geq 0 \ \ . $$ These conditions can be satisfied at some
$\vec{\xi_c}$. Let the number of $\vec{\xi_c}$ be equal to
$l_{0}$.

One can see that $B_b$ can be presented as $$ B_b = - \sum_i \xi_i
(\ln(n_i/n_{i\infty}) - \ln \xi_i - \ln f_i (\vec{\xi}) ) $$ The
term $-\ln(\xi_i)$ ensures that\footnote{Under the regular
behavior of $f_i$.} $$ \frac{\partial B_b }{ \partial
\xi_i}|_{\xi_i=0} = -\infty , $$ $$ \frac{\partial B_b }{ \partial
\xi_i}|_{\xi_i=1} = \infty  . $$ This leads to the following
properties of asymptotic behavior of $F_b$: \begin{itemize} \item
(A) The asymptotic thermodynamic channel can not be at $\xi_i = 1$
or $\xi_i=0$  precisely.
\item (B)
The previous property  doesn't
mean that $B_b$ goes to infinity, the equality $x \ln x |_{x
\rightarrow + 0 } \rightarrow  0 $ ensures that
at $\xi_i = 0;1$ the function $B_b$ attains some finite values.
 Here it is supposed that all
$f_i(\xi)$ have regular behavior.

\end{itemize}

The first property allows later to make some conclusions
about the properties of nucleation channels.
The second property establishes the possibility to
include the unary nucleation into the multicomponent
picture and to see the almost pure
channels of nucleation, which will be important here.

We shall denote coordinates of channels by $\xi_{as\ c}$. Consider
now the binary case, we enumerate the channels to have $$ \xi_{j\ as\ c\
1} < \xi_{j\ as\ c\ 2} < ... < \xi_{j\ as\ c\ l_{0}} $$

Every channel has the zone of interaction. The boundaries of
interaction zones are defined by relations
$$
\frac{\partial B_b}{ \partial \xi_i} = 0, \ \
\frac{ \partial^2B_b}{ \partial \xi_i^2} < 0
$$
The boundaries of these zones are straight lines in plane
$\vec{\xi}, \vartheta$. They are perpendicular to
$\vec{\xi}$-plane. We have to get the coordinates of these lines.
We shall denote these coordinates as $\xi_{j\ as\ b\ i}$ (the
first index marks component, the last is  the number, index "$b$" means
"boundary").   The total
number of these coordinates is $l_0 -1$. They are distributed as
$$
\xi_{j\ as\ b\ 1}
<
\xi_{j\ as\ b\ 2}
< ... <
\xi_{j\ as\ b\ l_{0}-1}
$$

One can see that
$$
\xi_{j\ as\ c\ i-1} < \xi_{j\ as\ b\ i-1} < \xi_{j\ as\ c\ i} , \ \
\xi_{j\ as\ b\ i-1} < \xi_{j\ as\ c\ i} < \xi_{j\ as\ b\ i}
$$

The boundaries of the zone of interaction for the first channel
are: the lower boundary is $\xi_{j\ as\ b\ 0} \equiv 0$, the upper
boundary is $\xi_{j\ as\ b\ 1}$. The boundaries of the zone of
interaction for the last channel are: the upper boundary is
$\xi_{j\ as\ b\ l+1} \equiv 1$, the lower boundary is $\xi_{j\ as\
b\ l_{0}-1}$. The boundaries of the zone of interaction for other
channels (number $i$) are: the upper boundary is $\xi_{j\ as\ b\
i} $, the lower boundary is $\xi_{j\ as\ b\ i-1}$.

 One can determine the time of relaxation for every
channel. It is necessary only to linearize the  system of
equations for the law of the embryos moving in plane $\nu_i$ near
the coordinates of the channel. Then we have the ordinary
differential equations with constant coefficients with known
solution as relaxation exponents\footnote{Here for the asymptotic
channels the relaxation resembles the relaxation in the kinetic
channel, which will be considered below. }.

One has to note that for the two component case this picture is
exact, but for multi component case the situation is more
complex: in plane  $\vec{\xi}$ together  with focuses
we have saddle points. The analysis in this case
here is the standard one for
the theory of dynamic systems.

The main ground for this
analysis is the reduction of $F$ into the product of
 intensive part $B$ and extensive
part $\vartheta$ ensures that  the thermodynamic channels  bottoms
(when $\vartheta$ is fixed) are straight  lines. The same is
true for the bounds of interaction zones - they are also straight
lines in plane $\vec{\xi}, \vartheta$, perpendicular to the
$\vec{\xi}$ plane.

Also one has to note that even
in the two component case there exist some illegal channels
among
channels of nucleation . They are
situated
in the region of the miscibility gap, which is defined by condition
 $B_b >0$.

One can give an interpretation of the function $B_b$ as
$\bar{\Delta \mu}$ - it is the mean value of chemical potential
excesses
$$
B_b = -
\bar{\Delta \mu} = -  \sum_j \xi_j  \Delta \mu_j
$$

An interesting question, which will have important consequences
is a question whether
it is possible to see the asymptotic thermodynamic channel
at some $\xi_i \approx 0$ or at least at very small $\xi_i$.
We shall these channels as the "pure" channels. The
statements $(A)$ and $(B)$ show that it is quite possible but
requires the specific behavior of activity coefficients $f_i$.
One can extract the following properties
 of such channels.
\begin{itemize} \item (C) Dependence of $f_i$ over $\vec{\xi}$
must be sharp.

It means that for some $\xi_j$ the inequality $ |
\partial  f_i/\partial \xi_j | \gg 1$ near the coordinate of the
pure channel $\xi_i \approx 0$  has to be satisfied.

\item (D) The pure channels are narrow and deep\footnote{ For the
narrow channel the characteristic variations of $\xi_j$ are many times
less than $1$.}.

The qualitative explanation is the following.
 The property $|\partial f_i/ \partial \xi_j | \gg 1$ means that near the
coordinate of channel bottom the absolute value  of $f_i$ ordinary
(not precise) attains big values. The existence of a channel
together with monotonous behavior of $\ln (\xi_j)$ at small
concentrations
 means that $f_i$ has a minimum
near the coordinate of a channel.  The channel becomes
deep.
The non-linear
behavior
of $\ln (\xi_i) $
at $\xi_i \approx 0$ ensures that the pure channel is
narrow.

\item (E) Positions (coordinates) of channels are rather
insensible to $n_i$.

Since $f_i$ has a deep minimum it means that
slight variations of $n_i$ can not lead to essential variation of
coordinates $\xi_{i\ as\ c}$ of a pure channel. \end{itemize}

The term "pure channel" doesn't mean that this channel corresponds
to the
absolutely  pure condensation of one component. The  entropy terms
$\ln(\xi_i)$ lead to existence of other components as species.
This will be the main source of specific kinetic behavior
discussed below.

It seems now that the situation with the pure channel is rather
rare in nature. Really, the behavior of $f_i$ is required to be
rather specific. But it is an illusion. Ordinary, in the situation
with immiscible components we have the mentioned behavior and the
pure channels of nucleation.

\subsection{Real thermodynamic channels}

Now we shall include into consideration the surface term $F_s = S
\sigma$, where $S$ is the surface of an embryo and $\sigma$ is the
surface tension. The free energy of an embryo formation will be $F
= F_b + F_s$. This form is the so-called capillary approximation. All
other correction terms are neglected here. The  term $F_s$ is
asymptotically negligible.

We shall try now to repeat all constructions for $F$ instead of
$F_b$. Really it is possible to do as far as $F$ can be presented
in the following form
$$
F = +  \Omega B(\vec{\xi}) + \Omega^{2/3}
\ \ ,  $$
where $\Omega = (S \sigma)^{3/2}$, then $$ B(\vec{\xi})
= \frac{B_b (\vec{\xi})}{\sigma^{3/2}(\vec{\xi}) 6 \pi^{1/2}
\bar{v}} $$ Here $\bar{v}$ is the mean value of the volume per one
molecule in a liquid phase defined as $$ \bar{v} = \sum_j v_{l\ j}
\xi_j $$ and $v_{l\ j}$ is the volume occupied by a molecule of
$j$-th component in a liquid phase. Then $$ F_c= \Omega^{2/3}/3 $$
 (index $c$ marks
the critical embryo) and the ordinary Gibbs rule takes place.

Later we suppose that there are no irregularities in behavior of
$f_i, v_{l\ i}$ and $\sigma$ as functions of $\vec{\xi}$.

Now we can see the maximum over the extensive variable $\Omega$
and again we see the set of channels defined by the extremums
 of function $B$. We can
reproduce here all arguments and constructions as we have done for
$F_b$. Again we observe properties (A)-(E). Certainly, the
numerical values of  coordinates of channels will be  another. We
shall call these channels as "thermodynamic channels of
nucleation".

The difference between consideration of the real thermodynamic
channel and the asymptotic thermodynamic channel is only in the
process of relaxation. For the asymptotic channel one can not
observe the saddle point in thermodynamic channel. So, there is no
use to speak about the quasi equilibrium along $\vartheta$ in the
precritical
region. For the real
thermodynamic channel one can speak about the quasi equilibrium.
Then one can estimate the time of relaxation $t_r$ over
$\vec{\xi}$ in a way it had been done by Zeldovitch for the near
critical region and come to
$$ t_r \sim \vartheta^{1/3}.
$$
 Also
one has to add to this time the time of regular motion to the
viscinity of the "bottom" of the channel, i.e. from initial values
to the region
$|\xi_i - \xi_{i\ c}| \sim \nu_{i\ c}^{-1/2}
|\partial^2 B / \partial \xi^2_i|^{-1/2} 2^{1/2}$. Since
$B$ is arbitrary, we prefer not to give the final expressions here,
because it can be done only on the base of the behavior of $B$
after the evident integration.

Here it is  necessary to outline the meaning of $\bar{v}$. One has
to keep in mind the Gibbs absorption equation or the Gibbs-Duhem
relations. In application to the embryo description it means the
phenomenological rule, which states that having  come   to
$\bar{v}$ and $\vec{\xi}$ one can simply  neglect all derivatives
of $\bar{v}$ over $\xi_i$. The same has to be done with
derivatives of $\sigma$ over $\xi_i$, which is known as the
so-called Wilhemsky-Renninger effect \cite{Renninger}. This rule,
which is very useful in practice, can be shown on the base of Gibbs
absorption equation.

In transition from $F_b$ to $F$  the surface tension have
appeared. Surface tension can have rather arbitrary behavior on
concentration. Can it violate the previous conclusions?  The
typical behavior of $\sigma$ as function of $\xi$ is the
following: as long as possible there exists an enrichment of a
surface layer and the surface tension doesn't variate essentially.
Certainly, the entropy of mixture has to be contained in $\sigma$.
Only when there is practically no molecules of component with a
small $\sigma$,
then the value of $\sigma$ begins to grow and attains
for the pure substance the value corresponding to $\sigma$ for
this pure substance. It means that the pure thermodynamic channel
of nucleation for component with maximal $\sigma$ is practically
impossible\footnote{ Certainly, the sharp dependence of $\sigma$
can be compensated by peculiar behavior of $f_i$.}.

This picture of the surface tension behavior has one important
qualitative  consequence. Precisely speaking the asymptotic
thermodynamic channels have nothing in common with precise
thermodynamic channels and are useless. Certainly, for the ideal
solution (all $f_i \equiv 1$) the additional entropy of mixture is
$$
(n_{surf}/(n_{surf}+n_{i\ surf})) \ln(n_{surf}/(n_{surf}+n_{i\
surf}))+
$$
$$
 (n_{i\ surf}/(n_{surf}+n_{i\ surf})) \ln(n_{i\
surf}/(n_{surf}+n_{i\ surf}))-
$$
$$
 (n_{bulk}/(n_{bulk}+n_{i\
bulk})) \ln(n_{bulk}/(n_{bulk}+n_{i\ bulk}))-
$$
$$
(n_{i\
bulk}/(n_{bulk}+n_{i\ bulk})) \ln(n_{i\ bulk}/(n_{bulk}+n_{i\
bulk}))
$$
where $n_{i\ surf}$ is the density of given component
in the surface layer, $n_{i\ bulk}$ - the same value for the bulk
phase, $n_{surf}$  - the density of all other components in a surface
layer, $n_{bulk}$ - the same value for the bulk
phase\footnote{Here we suppose that the function $\bar{v}
(\vec{\xi})$ is very smooth. }. Then the sum over all components has to
be taken. We see that the additional entropy of mixture at
moderate $\xi_i$ cannot be a sharp function of concentration.
Under the rough generalization one can speak about the flat region
- the region of the flat behavior of $\sigma$ as function of
$\xi_i$ and about the sharp region - the region of the sharp
dependence of $\sigma$ on $\xi_i$. In the sharp region the
thermodynamic channel of nucleation can be hardly situated,  then
this region isn't important. In the flat region the derivative of
surface tension is small and it means that the position of the
thermodynamic channel can not be essentially changed by inclusion
of the surface term. So, one can use the asymptotic (bulk)
thermodynamic channels as initial approximation to find the
coordinates of the real thermodynamic channels.

One has to show inclusion of the picture obtained in consideration
of $F_b$ into the picture in consideration of $F$ when $\vartheta$
goes to infinity. Precisely speaking the roots of $B$ don't
coincide with the roots of $B_b$. This is because the roots of $B$
are coordinates of channels when the extensive variable
$\Omega^{3/2}$ is fixed and the roots of $B_b$ are coordinates of
channels when $\vartheta$ is fixed. Certainly, these coordinates
will be different even when the channels would be the same
ones\footnote{ When $\Omega \rightarrow \infty$ or $\vartheta
\rightarrow \infty$ then $|F-F_b| \ll F$, but this doesn't mean
the similarity of coordinates of channels.}. To compare channels
one has to go from $\Omega$ to $\vartheta$ according to $$ \Omega
= \sigma^{3/2} 6 \pi^{1/2} \bar{v} \vartheta \ \ . $$ Then the
necessary inclusion will be evident.

\subsection{Kinetic channels of nucleation}

Now we shall turn to the definition of kinetic channels of nucleation.
Really, the mentioned thermodynamic channels of nucleation are
important for transition over the activation barrier, but when the
droplets are moving along the slope of the free energy profile
then kinetic terms are extremely important. At the sizes strongly
greater than the critical sizes one can not speak about the
absolute extremums and saddle points, which are independent from
kinetic factors. Now we shall investigate embryos with $\nu_{i\
c}$ more than three-four times greater than the corresponding
critical values.

 Now we have to investigate kinetic equations and
to determine their stationary points. The simplification here is
the possibility to use only the regular laws of
growth\footnote{This requires  a special justification for the
supercritical embryos
-
the embryos, which
contain much more molecules of every component than
the critical ones. This justification  can be
given analytically.} for embryos (since they are supercritical
embryos here they will be called as droplets).

The regular laws of motion
$$ \frac{d\nu_i}{dt} = W_i^+ - W_i^- \
\ ,
$$
where $W_i^+$ and $W_i^-$ are direct and inverse flows on
the embryo (the coefficients of condensation are included in these
flows) lead to expression for $$ \frac{d\xi_i}{dt} = - \xi_i
\vartheta^{-1} \sum_j \frac{d\nu_j}{dt} + \vartheta^{-1}
\frac{d\nu_i}{dt} \ \ . $$ So, the coordinate of a kinetic channel
$\xi_{i\ k}$ is determined\footnote{Here and later the lower index
$k$ denotes the stationary characteristics in the kinetic channel.
} by
$$ - \xi_i
 \sum_j \frac{d\nu_j}{dt} +
 \frac{d\nu_i}{dt} = 0 \ \ .
$$
It is necessary to stress that the last equations will be
equations on $\vec{\xi}$, the dependence on $\vartheta$ is absent
here. Really, one can write for $W_i^+$ the following formulas
from a simple gas kinetic theory (it is supposed that gas is an
ideal one and the regime of substance exchange is the free
molecular one) $$ W_i^+ = \frac{1}{4} v_{i\ t} S n_i \alpha_i $$
Here $v_{i\ t}$ is the mean thermal velocity of a molecule of
component $i$, $\alpha_i$ is condensation coefficient\footnote{It
can be included into the mean thermal velocities $v_{i\ t}$.}. The
surface square of an embryo $S$ is the same as in the expression
for the free energy, it differs only in microscopic corrections.
The kinetic coefficient $W_i^-$ can be obtained on the base of
$W_i^+$ by the
detailed balance  relation
$$
W_i^-(\vec{\nu}, \nu_i) =
W_i^+(\vec{\nu}, \nu_i-1) \exp(-F(\vec{\nu},\nu_i-1))
\exp(F(\vec{\nu},\nu_i))
$$
In the supercritical region near the
bottom of the channel the function $\exp(-F(\vec{\nu},\nu_i-1))
\exp(F(\vec{\nu},\nu_i))$ doesn't depend on $\vartheta$ when $V =
\sum_i v_i \nu_i$ is going to infinity and one can easily see that
${d\xi_i}/{dt}$ can be presented as $\vartheta$ multiplied by some
given function of $\vec{\xi}$. It allows to speak about the line
as a limit of  kinetic channel in a plane $\vartheta, \vec{\xi}$.
Now we shall determine this line concretely. Instead of the
previous relation one can write
$$
W_i^-(\vec{\nu}, \nu_i) =
W_i^+(\vec{\nu}, \nu_i-1) \exp(\partial F(\vec{\nu})/\partial
\nu_i)
$$
Calculation of $\partial F(\vec{\nu})/\partial \nu_i$
leads to
$$
\partial F(\vec{\nu})/\partial \nu_j =
-\ln(\frac{n_j}{n_{j\ \infty}\xi_j f_j(\vec{\xi})}) + (6
\pi^{1/2})^{2/3} \frac{2}{3} (\sum_i v_i \nu_i)^{-1/3} \sigma v_j
+ G.D.
$$
Here  $G.D.$ marks the terms, which according to the
Gibbs-Duhem relation have to be cancelled\footnote{One can argue
whether the Gibbs absorption relation can be applied to a growing
embryo, but we have to adopt this equation in order to have
concrete calculable results.}.

The Gibbs-Duhem equation for this case
can be written as
$$
\sum_{j'} \sum_j \nu_j \frac{-\partial
\ln(\frac{n_j}{n_{j\ \infty}\xi_j f_j(\vec{\xi})})} {\partial
\xi_{j'}} \frac{\partial \xi_{j'}}{\partial \nu_i} + \sum_{j'}
(6\pi^{1/2})^{2/3} \frac{\partial}{\partial \xi_{j'} } [\sigma
(\sum_j v_j \nu_j)^{2/3}] \frac{\partial \xi_{j'}}{\partial \nu_i}
=0
$$
for arbitrary $i$.
This Gibbs-Duhem relation isn't the standard one  because it
contains the
 term, which has a low  dimension. This equation appears as a
combination of the Gibbs-Duhem equation for the bulk phase and
the Gibbs-Duhem equation for the surface layer. One has to keep in
mind that due to the surface enrichment the concentration $\xi_i$
in a bulk solution
of an embryo can not be precisely defined as $\nu_i/
\vartheta$. This  note leads to the last relation.

One can see that the term $(6\pi^{1/2})^{2/3} \frac{2}{3} (\sum_i
v_i \nu_i)^{-1/3} \sigma v_j$ is negligible in comparison with
$\ln({n_j}/({n_{j\ \infty}\xi_j f_j(\vec{\xi})}))$ when the volume
of an embryo $ \sum_i v_i \nu_i$ is big enough. So, one can
neglect the correction term and come to
$$
dF(\vec{\nu})/d\nu_j
=
-\ln(\frac{n_j}{n_{j\ \infty}\xi_j f_j(\vec{\xi})})
$$
This relation is typical for asymptotic kinetic channels of
nucleation. Here one can already speak about the straight channels
of nucleation because the rhs doesn't contain extensive terms and
$d\nu_j/dt$ can be presented as $S$ multiplied by a function of
intensive terms.

The Gibbs-Duhem  relation becomes here also more standard and can
be written as\footnote{
Here appears a special question whether it is possible
to consider $\nu_i$ as $\vartheta \xi_i$ and to take
$\vartheta
\partial[\xi_j \ln(n_j/n_{j\infty} \xi_j f_j)]/\partial \xi_j$
instead of
$
\partial[\nu_j \ln(n_j/n_{j\infty} \xi_j f_j)]/\partial \xi_j$
in two previous Gibbs-Duhem relations. Since  the channels of nucleation
are rather
narrow ($\xi_i - \xi_{i\ c} \sim \vartheta^{-1/2}$) this can lead
only to correction terms with negligible order in powers of
$\vartheta$.
}
 $$ \sum_{j'} \sum_j \nu_j \frac{-\partial
\ln(\frac{n_j}{n_{j\ \infty}\xi_j f_j(\vec{\xi})})} {\partial
\xi_{j'}} \frac{\partial \xi_{j'}}{\partial \nu_i} =0
\ \ .
$$
This form
is absolutely typical for the bulk solution with standard
definition of concentration $\xi_i$ as $\nu_i/\vartheta$.

For asymptotic kinetic channel  the relaxation
to the stationary value of concentration
is practically evident. The
absence of curvature here  is
very important in contrast to the thermodynamic channels in the
under-critical or near-critical regions. The matter is that the
variable $\vartheta$ in the super-critical region has to be
regarded as the fast variable. This lies in contradiction with
consideration of the near-critical region (where $\xi_i$ are the
fast variables and $\vartheta $ is the slow one).

Consider the near-critical region.
 Having calculated $dF/d\Omega$ and $dB/d\xi_i$ one can
easily see that if there are no small parameters in $W_i^+, f_i$
and their derivatives, then the relaxation to the bottom of
thermodynamic channel takes place and the characteristic time of
this relaxation is estimated as $\sim  \vartheta^{1/2}/ W_i^+ $.
The time to overcome the barrier in the
near-critical region  has the estimate $\sim \vartheta
^{2/3}/W_i^+$ and   one can see that there exist the hierarchy
between these characteristic times, which provides that along the
stable variable $\sim (\xi_i-\xi_{i\ c})$ there is the quasi
equilibrium.

The opposite situation
takes place  in the super critical region. As we
shall see later the variable $\vartheta$ grows very rapidly, this
property will be called as the avalanche consumption
of metastable phase (see
\cite{TMP}).
 When the channel is straight one then
there is no difference whether $\vartheta$ is the fast
variable or not and  one  can see such a
relaxation. But  when the channel is curved there can be no
relaxation to the bottom  (the term "bottom" at the slope becomes
rather relative). So, here the asymptotic disappearance of
curvature plays a very important role. Really, the
velocity\footnote{When the embryo is moving near the stationary
value, i.e. the concentration alterations are small.} $d\vartheta / dt
$ allows the estimate\footnote{Here and later the lower index $k$
denotes the equilibrium value in kinetic channel.}
 $$
 d\vartheta / dt \sim \vartheta^{2/3} \rightarrow
\infty,
$$
$$
d(\xi_i - \xi_{i\ k})/dt \sim \vartheta^{-1} d\nu_i/dt \sim
\vartheta^{-1/3} \rightarrow 0 . $$
 Fortunately $\vartheta \sim
t^3$ and
$$
\int\frac{d(\xi_{i}-\xi_{i\ k})}{dt} dt \sim \int dt/t
\sim \ln t \sim \infty \ \ .
$$
 So, relaxation really takes place, but it
isn't the exponential, but the power  relaxation\footnote{All
details can be seen in \cite{preprint1}, Appendix 1 and Appendix
2.}. Here the straightness of the channel plays the principal
role.

We shall enumerate kinetic channels by the number of thermodynamic
channels, which this kinetic channel will be attained
from (So, the
kinetic channel can have several numbers or no number at all.
Kinetic channel can be described by one of its possible numbers.).

One can not investigate in details the transition from the
critical region to the region of sizes, which strongly exceed the
critical sizes. One of effective approximations is  the
$\delta$-function approximation suggested in \cite{preprint1}.
 Let some stationary value
of concentration $\xi_i$ (it is one dimensional variable for the
binary system)  be $\xi_{i\ +}$. The characteristic width $\Delta
\xi_i$ along $\xi_i$ allows the estimate
$$\Delta \xi_i \sim
\vartheta^{-1/2}\ . $$
The last value is relatively small. So, in the
law of growth one can put $\xi_i \approx \xi_{i\ +}$. Then
$$
\frac{d \vartheta}{dt} \sim  \vartheta^{2/3} \ \ ,
$$
which leads to
$\vartheta \sim t^3$. The characteristic deviation in
plane $\nu_i$ can be estimated as
 $(\xi_i-\xi_{i\ +}) \vartheta$
and has a characteristic value $\vartheta^{1/2}$. Having rewritten
the last estimate as a function of time we get $(\xi_i-\xi_{i\ +})
\vartheta \sim t^{3/2}$. We see that this width grows faster than
the width of
the diffusion blurring $\sim t^{1/2}$.
Then one can consider the pure diffusion over $\vec{\xi}$. The
solution of this problem is known.
The
distribution $n$ over $\vartheta, \vec{\xi}$ with $\delta$-like
initial conditions\footnote{Because
the  near-critical region has the
small relative sizes.} will be presented as
$$ n \sim
n_{\vartheta}\prod_i n_{\xi_i}, $$
where $n_{\vartheta} \sim \int
d\vec{\xi} n$  is simply a translation of initial
conditions (i.e. the boundary conditions at small\footnote{But
still greater than critical.}
$\vartheta$)\footnote{In the stationary conditions it is the
stationary rate of nucleation divided by the regular law of growth
$d\vartheta / dt$}
and  $n_{\xi_i}$
approximately satisfies  the diffusion equation
$$
\partial n_{\xi_i} / \partial t \sim D_i \partial^2 n_{\xi_i} /
\partial \xi_i^2
$$
with characteristic diffusion
 coefficient\footnote{All
$W^+_i$ are supposed here to have
 one and the same order.}  $D_i$
 and looks like the Green function of diffusion
on  the flat energy profile
$$
G \sim \exp(-(\xi_i-\xi_{i\
+})^2/(4D_it)) / (4 \pi D_i t )^{1/2}
$$
Approximation $D_i
\approx const$ as the function of size, which was used to get the
previous relation isn't too
suitable. Now we shall extract
coordinates with a real physical meaning.

 Having noticed that $\vartheta \sim t^3,
\partial/\partial t \sim \partial
/ \partial \vartheta^{1/3}$
  we can write instead the
previous diffusion equation the following equation
$$
\partial
n_{\xi_i} /
\partial \vartheta \sim D_{0 i}
\partial^2 n_{\xi_i} / \partial \xi_i^2\ .$$
 Here $D_{0 i}$ is the
new diffusion coefficient for evolution in $\vartheta, \vec{\xi} $
plane. The really significant variables are namely these
variables. A solution of the last equation has
a good approximation   also in the form of
the Green function
$$
G \sim
\exp(-(\xi_i-\xi_{i\ +})^2/(4D_{0i}\vartheta)) / (4 \pi D_{0i} \vartheta )^{1/2}
\ \ . $$

The diffusion over the plane surface can not lead to the
approaching of $\xi_i$ to $\xi_{i\ k}$. To see this approach one
has to take into consideration the terms corresponding to the
regular growth.

For $\xi_i$ near $\xi_{i\ k}$ one can restrict the regular law of
growth only by linear terms in $(\xi_i - \xi_{i\ k})$ series. For
diffusion
terms one can take the ordinary coefficient of
diffusion. So, we observe the evolution in a square potential.
Then one can prove that it is possible to use the Fokker-Planck
approximation.

The Fokker-Planck equation in multi dimensional harmonic potential
has a well known solution. Multidimensional square potential in
$\vec{\xi}$ is invariant over Lorenz transformation, which leads to
the separation of variables. We shall denote the new variables
also by $\xi_i$ here.

Since Lorenz transformation is a linear one, it can not lead to
any terms except linear ones along $\vartheta$. Now we have $(l-1)$
separate diffusion problems with known solutions (the Green function
is known, the eigenfunctions and eigenvalues of harmonic oscilator are
known, one can act in both approaches).

\subsection{The coordinates of channels}

Now we shall speak  mainly about asymptotic kinetic channels and
forget about the curved kinetic channels. The reasons will be the
following:
\begin{itemize}

\item When the channel is curved it means that the current
 characteristics aren't too far from the critical ones. Then the
 the derivatives
$\partial F /
\partial \nu_i$  are small ones and one has to take into account the diffusion.
The interesting situation is far from the critical region where the
channel is straight.

\item The characteristic size of relaxation to the stationary
value of concentration strongly exceeds the distance where the
curvature of a channel is essential.

\item
Later namely the characteristic sizes
corresponding to the
asymptotic channels will be important for
construction of the global kinetics of the process.

\end{itemize}

We shall mark the coordinates of asymptotic kinetic  channels also by the
subscript $k$. We shall call the asymptotic kinetic channels
simply as the kinetic channels.

 To determine this coordinate of equilibrium in the kinetic
channel one has to write the
laws of regular growth
$$
\frac{d\nu_i}{dt} = \frac{1}{4} S v_{i\ t} \alpha_i (n_i - n_{i\ \infty}
\xi_i f_i (\vec{\xi}) )
$$
at the supercritical asymptotics.

The surface square $S$ is given by $S= (6 \pi^{1/2})^{2/3}
\bar{v}^{2/3}(\vec{\xi}) \vartheta^{2/3}$,
which leads to
$$
\frac{d\xi_i}{dt} = \frac{(6\pi^{1/2})^{2/3}}{4 \vartheta^{1/3}}
\bar{v}^{2/3}
[
\alpha_i v_{i\ t} (n_i -n_{i\ \infty}
\xi_i f_i (\vec{\xi}) ) -
\xi_i \sum_j \alpha_j v_{j\ t} (n_j -n_{j\ \infty}
\xi_j f_j (\vec{\xi}) ) ]
$$
Here one can extract the function
$$
K_i =
[\alpha_i v_{i\ t} (n_i -n_{i\ \infty}
\xi_i f_i (\vec{\xi}) ) -
\xi_i \sum_j \alpha_j v_{j\ t} (n_j -n_{j\ \infty}
\xi_j f_j (\vec{\xi}) )]
$$
Equations $$K_i=0$$
 for all $i$ together with
 $$\sum_j \xi_j =1$$
form the system of nonlinear equations for coordinates\footnote{It
is necessary that the square form $\sum_i \sum_j (\partial^2 F/
\partial \xi_i \partial \xi_j )|_k (\xi_i - \xi_{i\ k}) (\xi_j -
\xi_{j\ k}) $ will be positively defined. Later only these
$\xi_{i\  k} $ will be considered.} $\xi_{i\ k}$. This system has
at least one solution when the vapor mixture is metastable.
It can be derived from the boundary behavior. But
sometimes this system can have several solutions (this corresponds
to several channels).

Near $\vec{\xi_k}$ one can rewrite the previous equation as
$$
\frac{d\xi_i}{dt} =\frac{(6\pi^{1/2})^{2/3}}{4 }
\vartheta^{-1/3} \frac{ d K_i(\vec{\xi}) }{d\xi_i}|_{\xi_i=\xi_{i\
k}} (\xi_i - \xi_{i\ k}) \bar{v}^{2/3}|_{\xi_i = \xi_{i\ k}}
$$

Equation for $\vartheta$ will be
$$
\frac{d\vartheta}{dt} =
\frac{(6\pi^{1/2})^{2/3}}{4}\vartheta^{2/3}
\bar{v}(\vec{\xi})^{2/3} \sum_i v_{i\ t} \alpha_i
(n_i - n_{i\ \infty} \xi_i f_i
(\vec{\xi}))
$$
or
$$
\frac{d\vartheta}{dt} =\vartheta^{2/3} P(\vec{\xi})
$$
where
$$
P(\vec{\xi}) =
\frac{(6\pi^{1/2})^{2/3}}{4}
\bar{v}(\vec{\xi})^{2/3} \sum_i v_{i\ t} \alpha_i
(n_i - n_{i\ \infty} \xi_i f_i
(\vec{\xi}))
$$

One can suggest the simple approximation when $\xi_i$  is near
$\xi_{i\ k}$ for all $i$ (here some given channel is already
chosen)
$$
\frac{d\vartheta}{dt} =\vartheta^{2/3} P(\vec{\xi_{i\ k}})
$$
This equation can be easily integrated
$$
\vartheta^{1/3} (t) = (t-t_0)P(\vec{\xi_{i\ k}})/3
$$
where $t_0$ is the time of appearance of a droplet in a
supercritical region\footnote{Initial size can be put to zero.}.

It allows to rewrite the equation for ${d\xi_i}/{dt} $ as
$$
\frac{d\xi_i}{dt} =\frac{(6\pi^{1/2})^{2/3}}{4 }
\frac{1}{
(t-t_0)P(\vec{\xi_{i\ k}})/3
} \frac{ d K_i(\vec{\xi}) }{d\xi_i}|_{\xi_i=\xi_{i\
k}} (\xi_i - \xi_{i\ k})\bar{v}^{2/3}|_{\xi_i = \xi_{i\ k}}
$$
with solution
$$
(\xi_i - \xi_{i\ k}) =  const (t-t_0)^L
$$
and unknown $const$ where
$$
L=\frac{(6\pi^{1/2})^{2/3}}{4 }
\frac{1}{
P(\vec{\xi_{i\ k}})/3
} \frac{ d K_i(\vec{\xi}) }{d \xi_i}|_{\xi_i=\xi_{i\
k}} \bar{v}^{2/3}|_{\xi_i = \xi_{i\ k}}
$$

If one can spread the last equation until $\xi_i= \xi_{i\ c}$
(index $c$ marks the critical value)
one can very approximately get
$$
const =(\xi_{i\ c} - \xi_{i\ k}) / (t_c-t_0)^L
$$
where
$\xi_{i\ c}$ is the concentration of the critical embryo (one has
to decide, which critical point has to be chosen), $t_c$ is the
time when according to the regular law of growth the embryo
attains the critical value of $\vartheta$.

One has to stress that there appear at least three problems, which
have to be solved
\begin{itemize}
\item
 1. how one can get the critical values of
$\vartheta$ and $\vec{\xi}$, i.e. $\vartheta_c$, $\xi_{i\ c}$;
\item
2. how one can get the asymptotic
 values of
$\vartheta$ and $\vec{\xi}$;
\item
3. how one can decide, which asymptotic channel corresponds to the
given critical point (i.e. to the given thermodynamic channel).
\end{itemize}
The first two problems are purely algebraic ones. One can only
note that to solve the first problem one has to use the Kelvin
relation  instead of direct search of minimum of function $B$
because of deviation of the bulk concentration from the integral
definition, which occurs due to the Gibbs absorption isotherm.

As an initial approximation for solution of the first problem one
can take the roots of the function $dB/d\xi_i$, where it
is necessary to forbid the differentiation of $\sigma$.
Then one can see the regions of interaction for all channels
having calculated $d^2 B/ d \xi_i d \xi_j$ and decided whether
the suspicious root is the bottom of the channel, the top of the
hill or the saddle point. Then
$dB/d\xi_i = 0$ leads to
$$
\bar{v} \frac{d \bar{\Delta \mu}}{d\xi_i}
=
 \bar{\Delta \mu} \frac{d \bar{v}}{d\xi_i} \  \ .
$$
This leads to
$$
\frac{\Delta \mu_i}{v_i} = \frac{\Delta \mu_j}{v_j} =  \frac{\bar{\Delta \mu}}{\bar{v}}
\ \ , $$
which is the known Kelvin relation.

Function $B_b$ can be found from $B$ when $v_i$ have approximately the same values
and $\sigma$ is independent on the concentration. sometimes the
search of the roots $d B_b / d \xi_i = 0 $ is more simple than
the
search of the roots $d B_b / d \xi_i = 0 $ and can be regarded as
some initial approximation.
When all $v_i$ have approximately the same values  then asymptotic
thermodynamic channels lies near the real thermodynamic channels.

To fix the positions of asymptotic kinetic channels of nucleation
sometimes
one can  use the approximation of strong supersaturations.
This approximation
means that in $K_i$  the value $n_{i\ \infty} \xi_i
f_i(\vec{\xi})$ is neglected in comparison with $n_i$. Then
$$
K_i = \alpha_i v_{i\ t} n_i - \xi_i \sum_j  \alpha_j v_{j\ t} n_j
$$
and together with $\sum_i \xi_i = 1$, (which now is trivial)
 we have the system of linear
equations with known solution
$$
\xi_i = \frac{\alpha_i v_{i\ t} n_i }{\sum_j  \alpha_j v_{j\ t} n_j }
$$
Here the solution is unique.

Another promising approximation is approximation of ideal
solution. In this approximation one can formally
split the system into two components: the first component and all
other components. Then one can get kinetic coefficients for these
two components and solve the two-component case where the equation
of the concentration of the first component will be the following
(obtained by D.Stauffer in \cite{stauffer})
$$
\alpha_1 v_{1\ t} (1-\xi_1) (n_1 - n_{1\ \infty}\xi_1) =
\alpha_2 v_{2\ t} \xi_1 (n_2 - n_{2\ \infty}(1-\xi_1))
$$
(it has two solutions: one for the bottom
(stable stationary value) of the channel and one
for the
top (unstable stationary value),
also it is necessary to check boundary values at
$\xi_1 = 0;1$ as possible channels). The problem is that $n_{2\
\infty}$ is unknown because the composition of the second complex
component is unknown, but now for this component the number of
substances is reduced and we can continue this procedure having
chosen in the second component two composing components, etc. As a
result we shall get the equation for concentration.
This completes the method of approximation of ideal mixture.

In the general case one can speak about invariant
$$
\alpha_i v_{i\ t} (n_i - n_{i\ \infty} \xi_i f_i) / \xi_i  = inv
$$
and
 write equation on this value. This way was described in
\cite{book}.

The third problem is rather hard to solve but one can note some
essential properties:
\begin{itemize}

\item

Ordinary one can use the regular  velocities of droplets growth
and ignore fluctuation terms. Only under a very specific behavior
of properties the fluctuation terms are essential. This can take
place only when there explicitly exist some  small parameters.

\item

Rather often it is possible to act in a following simple manner:
It is sufficient to get the regions of interactions for kinetic
channels and then to see to the region of what
kinetic channel belongs
the given thermodynamic channel. This way is very effective for
practical needs. The probability to make an error is very low.

\item

In any case it is impossible that droplets can cross the
miscibility gap. Results presented in \cite{djikaiev2} are wrong.
Really, in the miscibility gap the slope is positive (with
increase of $\vartheta$ the value of free energy also increases).
So here the kinetic channels will be
thermodynamic ones (if we don't take into account the disappearance of
the embryo), which have no
saddle points.
\end{itemize}

\section{Standard nucleation}

Nucleation is a period of a rather intensive formation of
droplets. To begin the investigation of the nucleation process one
can note that the number of thermodynamic channels have nothing in
common with the number of compositions of solution in droplets,
i.e. with the number of kinetic channels. The illusion of such
connection was presented in \cite{djikaiev2}, but there  are no
reasons to justify this illusion.

According to the
ordinary adopted classification one can extract in the
 condensation the first period - the period of intensive
formation of droplets and the second period - the period
after the nucleation but before the coalescence.
But it is necessary to stress
that even at the first period the variations of the vapor densities
$n_i$ or of activities $\zeta_i$ are rather important. This is in
contradiction with the statement made in \cite{djikaiev2} (page
391) where it is noted that $n_i$ during the period of nucleation
are constant. Under such assumption it is impossible to get any
spectrum of the droplets sizes\footnote{Here and later the size of
the droplet will be the characteristic, which grows with the
velocity independent of this characteristic. For the free
molecular regime such characteristic will be  $\nu_i^{1/3}$. The
concentrations here are  fixed.} different from the constant
amplitude. Nevertheless in \cite{djikaiev2} some attempts to get
the spectrum of sizes were made without any citations of the already presented
theory in
\cite{preprint0}, \cite{physicaa94}, \cite{preprint1}, \cite{preprint2}.
They follow instructions of
\cite{preprint0}, \cite{physicaa94}, \cite{preprint1},
\cite{preprint2} to consider the one dimensional equation instead
of  the system of condensation equations, which also were presented
in \cite{preprint0}, \cite{physicaa94}, \cite{preprint1}, \cite{preprint2}.

As it was noticed earlier $\xi_{i} \approx \xi_{i\ k}$. It means
that the quantity of the molecules of component $i$ in droplets
(i.e. $g_i$) satisfies equality
$$
g_i/\xi_{i\ k} = g_j/ \xi_{j\ k}
$$
for two different components.  This takes place when the
characteristic size of droplets during the nucleation period (the
first period) strongly exceeds the critical size and one can
assume that the evolution is mainly determined by rather big
droplets, which have $\xi_i \approx \xi_{i\ k}$. The process of
relaxation was described above.

Namely this relation is the base for reduction of multi-component
case to the one component condensation.
The relaxation processes haven't been
mentioned in \cite{djikaiev2} at all.
Really, having marked the
initial densities of vapor as $n_i(t=0)$ and the current
densities as $n_i (t)$ ($t$ is time) one can easily see that
$$
(n_i(0) - n_i(t))/\xi_{i\ k} =(n_j(0) - n_j(t))/\xi_{j\ k}
$$
for two different components $i$ and $j$. This forms the base for
reduction to the one component case.

We see that this reduction isn't the consequence of some
functional approximations used in \cite{preprint0}, \cite{preprint1}, \cite{physicaa94} and
repeated in \cite{djikaiev2}. So, it is more than symbolic that
to get the reduction to the one dimensional case
\cite{djikaiev2} also uses the functional
exponential approximation and approximation of a regular growth
(they are also from \cite{preprint0}, \cite{preprint1}), \cite{physicaa94},
which isn't necessary here and is the little
defect of narration\footnote{It this derivation it is supposed that the
equilibrium concentrations do not change too fast, which has to be
proven (and can be done).}
of \cite{preprint0}, \cite{preprint1}, \cite{physicaa94}.

 One can directly show that the droplets of big sizes are the
main consumers of vapor phase. One has specially stress attention
on this property despite it was considered in \cite{physicaa94}
because in \cite{djikaiev2} in was announced (p.391) that "this
hardly influences the results of our theory but significantly
simplifies the mathematical expressions". The cited conclusion is
wrong\footnote{It isn't clear how after this conclusion in
\cite{djikaiev2} it was possible to consider only the droplets of
big sizes without the logical destruction of the theory.}. This
property can be shown by the set of simple qualitative estimates.
We omit
here these estimates because this property will be directly
seen from the final results. Also it can be directly shown that
the quasi-stationary approximation for the droplets, which sizes
$\vartheta$ two or three times exceed the critical size
$\vartheta_c$ really takes place  (the concentrations here are
arbitrary, but
in reality
the quasi-stationary distribution is essential only
for concentrations near the stationary value in the given
channel).

Both these properties are based on the possibility of the
thermodynamic description of the critical embryo. In
\cite{djikaiev2} the possibility of analytical derivation of these properties
hasn't been mentioned at all.

On the base of the last statements one can write the balance
equation in the following form
$$
n_i (0) = n_i (t) + g_i(t)
$$
where $g_i(t)$ is the number of molecules in the liquid phase and
it can be presented as
$$
g_i (t) = \int_0^t dt' p_i(t') \hat{\nu_i} (t,t') dt'
$$
Here $\hat{\nu_i} (t,t')$ is the number of molecules inside the
droplets born at $t'$, $p_i(t')$ is the intensity of formation of
droplets. Our next task is to determine this intensity.

The intensity of the droplets formation depends on the height of
the activation barrier of nucleation. Since there are several
saddle points it seems that there are several possibilities for
embryos to become the supercritical objects of the liquid phase.
Ordinary, the leading channel is the unique one.
The intensity of formation is proportional to $\exp(-F_{c\ j})$
(here $F_{c\ j}$ is the height of activation barrier in the $j$-th
thermodynamic channel). It follows from the standard classical
consideration (see \cite{class}) in stationary conditions because
the nucleation rate can be calculated in stationary approximation.
Here it is sufficient to keep only the leading term and not to
think  about the preexponential factor, which is the matter of
numerous discussions \cite{stauffer}, \cite{tech}. It is important
that the normalizing factor of the equilibrium distribution for
every thermodynamic channel is one and the same
since all channels
go from the origin of coordinates.

 One can choose
the channel with the lowest $F_{c\ j}$. Let it be the first one.
If there is another channel with $| F_{c\ j} - F_{c\ 1}| \leq 1$
the theory presented below has to be slightly modified\footnote{
We have simply to add this intensity to the intensity in the first channel.
Functional dependencies on the vapor densities will be approximately the same.
One has to take into account the intensity with the smallest $\nu_{i\ c}$ (see later). }. But the
probability of such coincidence is very low since $F_{c\ j} \gg
1$. All these estimates have to be done at initial values $n_i =
n_i(t=0)$ when there is no depletion.  At first we shall
investigate the period of intensive formation of droplets in the
first channel. This period doesn't cover all nucleation
possibilities but as we shall see later namely this period will be
the main in nucleation.

From the first point of view it seems that the variations of $n_i$
can lead to the variations of $\Delta F_{i\ c}$ and sometimes the
channel, which was the first one can be changed by some other
channel, which becomes  the most energetically profitable one.
Fortunately the leading channel remains the main one at least
until the end of intensive nucleation in this channel\footnote{The
theory for this exclusion will be presented later.}.

The droplets consumes vapor and this effect strongly diminishes
the intensity of nucleation rate. To see this one can simply
calculate the derivatives of the free energy of the critical
embryo\footnote{The thermal effects are supposed to be absent as it has been noted earlier.}
$$
\frac{dF_c}{d n_i} =  - \frac{1}{n_i} \nu_{i\ c}
$$
The last relation can be seen directly from the explicit expression
for the free energy $F(\vec{\nu})$ if we notice that in
$$
\frac{dF_c}{d n_i} = \frac{\partial F}{\partial n_i } + \sum_j
\frac{\partial F}{\partial \nu_j}\frac{\partial \nu_j}{\partial n_i}
$$
the term $\sum_j \frac{\partial F}{\partial \nu_j}\frac{\partial
\nu_j}{\partial n_i} $ vanishes as far as $ \frac{\partial
F}{\partial \nu_j} =0 $ at the point of extremum (i.e. at the
saddle  point). This equality is a concrete realization of
nucleation theorem in the multi-component nucleation (see
\cite{preprint1}, \cite{preprint2} for the first derivation in the
application to the multicomponent condensation kinetics, but
generally this result was mentioned in \cite{hill}).

This relation plays the leading role in further analysis. To apply
the thermodynamic description of the critical embryo it is
necessary to have $\vartheta \gg 1$. Without the thermodynamic
description it is nothing to do in the quantitative theory of
nucleation. To use the approximation of solution in the bulk of
the embryo it is necessary to be $\nu_{i\ c} \gg 1$. Otherwise we
have to construct the model of liquid solution in the force field
of several molecules of rare  components. This problem is very
hard to solve. Then it is treated as heterogeneous condensation
and the molecules of rare components are considered as
heterogeneous centers. The theory of heterogeneous condensation has
 been given in \cite{heterog}. So, here
we have to suppose that all $\nu_{i\ c} \gg
1$ (but this doesn't mean that  $\xi_i$  essentially differs from
zero for the supercritical embryos).

On the base of $ - n_i dF_c/d n_i = \nu_i \gg 1$ one can easily see that
even the small fall of the molecular number density leads to the
interruption of intensive (in comparison with initial) formation
of droplets. It is the base for further approximations.

One can see that this point of view  radically differs from
opinion presented in \cite{djikaiev2} where inequalities like
$dF_c/dn_i \gg 1$ are considered (see \cite{djikaiev2} eq. (23) )
as some special artificial external restrictions.

Since $ n_i dF_{j\ c}/dn_i \gg 1$ then the main dependence of the
stationary intensity of formation $J_{s\ i}$ of the droplets in
the $i$-th thermodynamic channel is given by
\begin{equation} \label{app}
J_{s\ 1} (\vec{n(0)})
=J_{s\ 1} (\vec{n(t)})
\exp(-F_{1\ c}(\vec{n(0)}) + F_{1\ c}(\vec{n(t)}))
\end{equation}
This formula is for the first channel. It is valid
when
\begin{equation} \label{pp}
|n_i(t) - n_i(0)| \leq n_i(0) / \nu_{i\ c\ 1} |_{n_i =n_i(0)}
\end{equation}
This region of $n_i$ corresponds to the intensive formation of the
droplets through the first channel. The last estimate restricts
variation of $n_i$ and the process of formation takes place in
time. One can imagine that  a long tail with a small intensity  can
be more important than the short initial stage with a high
intensity. Then the last estimate fails. Fortunately it is easy to
see that every droplet grows with a growing intensity in time.
Then $ d^2 n_i /dt^2 < 0 $ and the estimate (\ref{pp}) really
corresponds to the region essential for nucleation.

One can see one very important feature for $n_i$ satisfying
(\ref{pp}):
\begin{itemize}
\item For $n_i$ satisfying (\ref{pp}) all values of $\xi_{i\ c\
1}$ are approximately constant\footnote{The last index is the number of a channel.}
(the relative variations of
$\xi_{i\ c\ 1}$ and $1-\xi_{i\ c\ 1}$ are small). For all other
thermodynamic channels the relative variations of equilibrium
concentrations are small if the channel is the "clear"
one\footnote{The "clear channel" is the channel with relatively
high sides (more than several thermal units). }.
\end{itemize}
The analogous property takes place also for kinetic channels
\begin{itemize}
\item For $n_i$ satisfying (\ref{pp}) all values of $\xi_{i\ k \ 1
}$ are approximately constant (the relative variations of $\xi_{i\
k\ 1}$ and $1-\xi_{i\ k\ 1}$ are small). For all other kinetic
channels the relative variations of equilibrium concentrations are
small if the channel is the "clear" one\footnote{Here the sides of
the channel are determined $\vartheta$ being fixed.}.
\end{itemize}

These properties can be justified analytically.

The estimate (\ref{pp}) also allows to make decomposition in the
argument of $\exp(-F_{1\ c}(\vec{n_i(0)}) + F_{1\
c}(\vec{n_i(t)}))$ in Tailor series. It is necessary to decide
what number of derivatives has to be taken to ensure  the high
accuracy. In \cite{djikaiev2} the parameters of analogous
decompositions were chosen in a wrong way. Really, in
\cite{djikaiev2} the correction term has an order proportional to
the first derivative $dF_c/dn_i$, the second derivatives were
ignored. It is wrong and has a hidden supposition that the second
derivative have the same order as the first derivative. This
supposition is too strong. Instead of this one has simply to
require that
$$
-F_{1\ c}(\vec{n_i(0)}) + F_{1\ c}(\vec{n_i(t)})
\approx
\sum_i \frac{dF_{1\ c}}{dn_i}|_{n_i = n_i(0)} (n_i(t) - n_i(0))
$$
The last approximation has to be directly checked for every type
of multicomponent mixture. It can be shown on
the base of the evident
inequalities
$$
|
-F_{1\ c}(\vec{n_i(0)}) + F_{1\ c}(\vec{n_i(t)})
|
 >
|
\sum_i
\frac{dF_{1\ c}}{dn_i}|_{n_i = n_i(0)} (n_i(t) - n_i(0))
|
$$
and
$$ |
-F_{1\ c}(\vec{n_i(0)}) + F_{1\ c}(\vec{n_i(t)})
| <
| \sum_i
\frac{dF_{1\ c}}{dn_i}|_{n_i = n_i(0)(1-1/\frac{d F_{1\ c}}
{d n_i}|_{n_i=n_i(0)})} (n_i(t) -
n_i(0))
|
$$

 There is no special parameter of
decomposition, the considerations  made in $O$-symbols presented in
\cite{djikaiev2} are irrelevant because in reality the values of
$dF_{1\ c}/dn_i$ are big,  but still finite. Nevertheless it is
possible to give some convincing arguments to see that the last
approximation is valid. We shall show that the terms with second
derivatives are small. For one component condensation the
required
property is obvious and can be checked by explicit formulas. In
the multicomponent nucleation since there are generally unknown
functions $f_i$ we can not get  the general
results. To estimate the second derivatives we have to consider
the values $(d\nu_{i\ c}/dn_j)/\nu_{i\ c}$.
They take moderate values, at
least they are smaller\footnote{Here it is supposed that all $n_j$
have one and the same order.} than $\nu_{i\ c}$. To see this property
one can use variables $\vartheta, \vec{\xi}$  and if  all $f_i$
are
directly
independent on $n_j$ one can see that $(d\nu_{i\ c}/dn_j)/\nu_{i\ c}$
has no small or
big parameters. So, $(d\nu_{i\ c}/dn_j)/\nu_{i\ c}$ takes
moderate values    less than
the values $\nu_{i\ c}$.

According to (\ref{pp}) in the law of growth one can take $n_i =
n_i(t=0)$ and get
$$
\frac{d\nu_i^{1/3}}{dt} =
\frac{1}{12} \alpha_i
 (6 \pi^{1/2})^{2/3} \bar{v}^{2/3}    \xi_i^{-2/3} v_{i\ t}
 (n_i(0) - n_{i\ \infty} \xi_i f_i (\vec{\xi}))
$$
The values of concentrations $\vec{\xi}$ have to be taken as the
equilibrium values in kinetic channel corresponding to the first
thermodynamic channel. Here we shall take  more strong definition
of the supercritical embryos. We shall call the supercritical
embryos as those embryos, which are already passed the critical
point and have already approached to the kinetic channel.

One has to check that even with a new definition the statement
that the supercritical droplets are the main consumers of vapor
remain valid. Under the conditions of applicability of
thermodynamic description of the critical embryo this statement
can be proven analytically with some reasonable assumptions. Also
the second statement about the quasistationary value of the
nucleation rate during the nucleation period has to be proved for
a new definition of the supercritical droplets. It also can be
done analytically in conditions of thermodynamic description of
the critical embryo.

One can  state that the same property as for thermodynamic channels
can be seen for kinetic channels. Namely, for variations of $n_i$
satisfying (\ref{pp}) the equilibrium concentrations of kinetic
channel can be considered as constant ones. This property is valid
for the "clear" kinetic channels
when that the derivatives
$d\xi_i / dt $ have no small
parameters in the
vicinity of the equilibrium values in kinetic channel.

For densities $n_i$ satisfying to (\ref{pp}) one can take also
$\vec{\xi}$ as constant values $\vec{\xi_{1\ k}}$ and get the rhs
as a constant value.  Then it is possible to integrate the last
equation and get
$$
\nu_i^{1/3} = \Lambda_i (t-t')
$$
where
$$
\Lambda_i = \alpha_i \frac{1}{12}
 (6 \pi^{1/2})^{2/3} \bar{v}^{2/3}    \xi_{i\ 1\ k}^{-2/3} v_{i\ t}
 (n_i(0) - n_{i\ \infty} \xi_{i\ 1\ k} f_i (\vec{\xi_{1\ k}}))
$$
Here it is taken into account the initial condition $\nu(t=t') =
0$

It is easy to see that
$$
\frac{\Lambda_i^3}{\xi_{i\ 1\ k}} =\frac{\Lambda_j^3}{\xi_{j\ 1\ k}}
$$
for different components.

The equation for $g_i$ will be the following
$$
g_i = \int_0^t dt' J(t') \Lambda_i^3 (t-t')^3
$$
Here $J(t')$ is the rate of nucleation.

For $J$ one can take according to the second statement the
following expression
$$
J \approx J_{s\ 1}
$$
where $J_{s\ 1}$ is the stationary rate of nucleation through the
first thermodynamic channel.

Then approximation (\ref{app}) leads to
$$
J = J_* \exp( -
\sum_i \frac{dF_{1\ c}}{dn_i}|_{n_i = n_i(0)} (n_i(t) - n_i(0))
)
$$
where
$$
J_* =
J_{s\ 1} (\vec{n}=\vec{n}(t=0))
$$
Then
$$
g_j = \int_0^t dt'  J_* \exp( -
\sum_i \frac{dF_{1\ c}}{dn_i}|_{n_i = n_i(0)} (n_i(t') - n_i(0))
)\Lambda_j^3 (t-t')^3
$$

Since
$$
n_i(t') - n_i(0) =  - g_i (t')
$$
we have
$$
g_j = \int_0^t dt'  J_* \exp(
\sum_i \frac{dF_{1\ c}}{dn_i}|_{n_i = n_i(0)} g_i(t')
)\Lambda_j^3 (t-t')^3
$$
and
$$
g_j = \int_0^t dt'  J_* \exp(
\sum_i \frac{dF_{1\ c}}{dn_i}|_{n_i = n_i(0)} \xi_{i\  k}
g_j(t')/\xi_{j\  k}
)\Lambda_j^3 (t-t')^3
$$
Then the last equation is the one dimensional equation with
solution, which describes the nucleation period. This equation is
the same as the in the situation of one component nucleation. For
the first time this fact was noted in \cite{preprint0}, then it
was confirmed in \cite{preprint1}, \cite{preprint2},
\cite{physicaa94}. So, there are no problems in solution and  in
the description of the nucleation period. In \cite{djikaiev2} this
known fact was presented as a new result. We need to discuss it
even in the one dimensional case  because the errors from \cite{4}
were accurately reproduced in \cite{djikaiev2}, which
points that
these errors were made not occasionally but were the  consequence
of deep misunderstanding of the nucleation kinetics. So, these
errors have to be corrected explicitly\footnote{Solution of the
one-dimensional equation in a more general case was presented in
\cite{preprint1}.}.

The last equation can be rewritten as
$$
g_j = W \int_0^t dt'   \exp( -
P
g_j(t'))
 (t-t')^3
$$
with
constants
$$
W = J_*\Lambda_j^3
$$
$$
P= -
\sum_i \frac{dF_{1\ c}}{dn_i}|_{n_i = n_i(0)} \xi_{i\  k}/\xi_{j\
k} > 0
$$

The physical  sense of renormalizations made in \cite{djikaiev2}
is wrong, the mentioned sense announced in \cite{djikaiev2} would
be correct if all densities $n_i$ are constant  up to the end of
the all surplus vapor depletion (due to the vapor depletion these
densities will decrease). Actually there is no need to fulfill
renormalizations with a physical sense, it is sufficient to write
equation for $G_j= Pg_j(t')$ as
$$
G_j = WP \int_0^t dt'   \exp(
 - G_j(t'))
 (t-t')^m ;
\ \ m=3
$$
The scale renormalization $t \rightarrow (WP)^{1/4} t$ leads to
$$
G_j =  \int_0^t dt'   \exp(
 - G_j(t'))
 (t-t')^m ;  \ \ m=3
$$

One can prove that this equation has the unique solution and one
can get this universal solution numerically. This approach was
outlined in \cite{heterog} where all details can be found. The
base of application is the rapid decrease of $J(t)$ after the end
of nucleation period, which allows to present the influence of
$g_i$
after  the end of nucleation via the first three moments
 of the distribution, i.e. via
 $\int_0^{\infty}  t'^i J(t') dt' , i=0,1,2,3$
  (including the zero moment, which is
the leading term).

The iteration solution with the zero approximation $G_i = 0$ and
the
recurrent procedure
$$
G_j^{(i+1)} =  \int_0^t dt'   \exp( -
G_j^{(i)}(t'))
 (t-t')^m \ \ , \ \ m=3
$$
can be also applied. But it is necessary to stress that the reason
of the rapid convergence of iterations is the big power $m=3$ in the
integral term. This fact was noticed for the first time in
\cite{Novos}. This property is the concrete realization of the
principle of the avalanche consumption formulated in \cite{TMP}.
Nevertheless in \cite{djikaiev2} it was stated that already the
first iteration gives the appropriate result for all positive
powers $m$ instead of $3$ in the integral equation. This
conclusion is wrong, which can be seen from the explicit solution
at $m\rightarrow + 0$, namely
$$
G_i = \ln(t+1)
$$
This function strongly differs from the first iteration
$$
G_i^{(1)} = t \ \ ,
$$
which is the correct form of the first iteration  for $m\rightarrow +
0$,
and also differs from
$$
G_i^{(1)} = t^4/4 \ \ ,
$$
which was used in \cite{djikaiev2}.
 The most surprising fact is that the main
part of \cite{djikaiev2} is devoted to get the estimates of the
accuracy of the first iteration and the
"necessary"
 estimate has
been proven. Certainly,  this estimate is wrong. We needn't to
follow procedure \cite{djikaiev2} but can simply compare the
universal\footnote{
For approximation of the universal solution
see \cite{PhysrevE94}. }
solution $G_i$ and the first iteration $G_i^{(1)}$. Both
curves are  $J= \exp(-G_i)$ and $J_1 = \exp(-G_i^{(1)})$ are drawn
in Fig.1. and they can be hardly separated. It is very simple to
show this approximate coincidence analytically.

\begin{figure}



Figure 1. The form of the size spectrum and the first iteration in dimensionless units.

\end{figure}

The description of the nucleation period is completed, but the
period of consumption of the main quantity of surplus substance
has to be described. This period is the main in the phase transition because
namely during this period the main quantity of surplus substance
is going from the mother phase into a new liquid phase.
Really, at the end of the nucleation
period the relative quantity of the substance is small.
It can be estimated as
$$
g_i = \Lambda_i^3 J_* \frac{t_0^4}{4}
$$
where the duration $t_0$ of the nucleation period can be estimated
as
$$
t_0 = (\frac{1}{4}  \sum_i |\frac{\partial F_c}{\partial n_i }|
\xi_{i\ k}
\frac{\Lambda_j^3 J_*}{\xi_{j\ k}})^{-1/4}
$$
(here the  first channel is considered).

This estimate is based on the analytical form of the first
iteration and it is clear that this estimate is too
long. At
least the values  $g_i$ satisfy  inequality
$$
g_i < |\frac{\partial F_c}{\partial n_i}|^{-1} = \frac{n_i}{\nu_{i
\ c }}
\ \
$$
(the critical value is referred for the first channel),
which shows that the relative quantity of substance in droplets is
small at the end of the nucleation period.

\subsection{Standard evolution after the end of nucleation}

Now we shall describe  the period of consumption of practically
all surplus  substance from the mother phase.  Precise solution of
the evolution during this period is unknown, the solution
presented in \cite{djikaiev2} is wrong. This solution artificially
without any comments
and justifications
prolongs some relations established at the
nucleation period.
This prolongation isn't correct.
Here we shall present some effective
approximate solutions of kinetic equations. Here we have no
necessity to ensure such high precision as at the nucleation
period because  there is no further nucleation (except the
special case, which will be described in the final part). So,
simply the high relative accuracy in $n_i$ is necessary.

 At first we shall use the property that at the
end of nucleation the relative quantity of surplus substance is
small. Then it is possible to use the monodisperse approximation
for $g_i$. This approximation is the following
$$
g_i = N \nu_i
$$
where
$N$ is the number of droplets (It can be taken as the result of
the nucleation
period
description) $N = \int_0^{\infty} J_{s\ 1} (t') dt' \approx
J_* t_0 0.9$.
Here $\nu_i$ is the coordinate of the monodisperse spectrum.
When  the monodisperce approximation for $g_i$ is bad, then
$g_i$ is negligile in the substance balance.

It is
necessary now to write kinetic equations for the time evolution
of $\nu_i$.  These equations can be written in approximation
of the regular growth. Then
\begin{equation} \label{w}
\frac{d\nu_i}{dt} =
\frac{1}{4}
v_{i\ t}  S [ n_i (t=0) - N \nu_i - n_{i\ \infty} \xi_i
f_i(\vec{\xi}) ]
\end{equation}
$$
S= (36 \pi)^{1/3} V^{2/3}, \ \ V  = \sum_j v_i \nu_i = \bar{v}
\vartheta
$$
Here and later $\alpha_i$ are included into $v_{i\ t}$.
 Then
\begin{equation}\label{w+}
\frac{d V^{1/3}}{dt} =
 \frac{1}{12}
  (36 \pi)^{1/3}
  \sum_i v_i v_{i\ t}
   [ n_i (t=0) - N \nu_i - n_{i\ \infty} \xi_i
f_i(\vec{\xi}) ]
\end{equation}

Here we conserve $\nu_i$ in $ [ n_i (t=0) -
N \nu_i - n_{i\ \infty} \xi_i
f_i(\vec{\xi}) ] $, which can lead to
 perspective  approaches when for $\nu_{i}$
some approximations are suggested\footnote{For
example, for fast component
one can put $\nu_i$ to its final value (see later).}.
One can rewrite the last expression as
\begin{equation} \label{rr}
\frac{d V^{1/3}}{dt} =
 \frac{1}{12}
  (36 \pi)^{1/3}
  \sum_i v_i v_{i\ t}
   [ n_i (t=0) - N \xi_i \frac{V}{\bar{v}(\vec{\xi})} - n_{i\ \infty} \xi_i
f_i(\vec{\xi}) ]
\end{equation}

For $\vec{\xi}$ the evolution equation will be the following
\begin{eqnarray}\label{xxi}
\frac{d\xi_i}{dt}
=
\frac{\bar{v}(\vec{\xi})}{V^{1/3}}
\frac{1}{4} (36 \pi)^{1/3}
[ (n_i(0) - N \xi_i \frac{V}{\bar{v}} -
n_{i\ \infty} \xi_i f_i ) v_{t\ i} -
\nonumber
\\
\\
\nonumber
\xi_i \sum_j
(n_j(0) - N \xi_j \frac{V}{\bar{v}} -
n_{j\ \infty} \xi_j f_j ) v_{t\ j}
]
\end{eqnarray}

It is clear that  one can not not solve two last equations
analytically (at least for the arbitrary coefficients of
activity). The analytical solution announced in
\cite{djikaiev2} as the precise one is wrong. Now we shall
suggest different approximations, which allow to solve these
equations.

Some useful  approximations  can be found  in \cite{preprint1},
\cite{preprint2} where it is shown that they can be presented as
the iteration procedures with
  rather
high rate of convergence.

{ \bf Approximation of ideal surface}

 In (\ref{w}) we shall consider $S$ as known function of time.
 Suppose that
$S$ is known in some approximation. Then one can see the following
iterative procedure
\begin{equation}\label{wig}
\frac{d\nu_i^{(k+1)}}{dt} =
\frac{1}{4}
v_{i\ t}  S [ n_i (t=0) - N \nu_i^{(k+1)} - n_{i\ \infty} \xi_i
f_i(\vec{\xi}) ]
\end{equation}
$$
S= (36 \pi)^{1/3} (\sum_j v_j \nu_j^{(k)})^{2/3}
$$
For initial $\nu_j^{(0)}$ one takes    $\nu_j^{(0)} =
\Lambda_j^3 t^3$.
One can  take in (\ref{wig})
$\xi_i$ as
$$
\xi_i = \frac{\nu_i^{(k)}}{\sum_j \nu_j^{(k)}}
$$
or
$$
\xi_i = \frac{\nu_i^{(k+1)}}{\sum_j \nu_j^{(k+1)}}
$$
The first opportunity  leads to the linear differential equations,
which can be solved and  it completes the iteration step.

{ \bf Approximation of the initial concentration }

Consider (\ref{rr})
and put in  this equation all $\xi_i$ to their initial values
$\xi_i(0)$,
which are known from the consideration of the nucleation period.
Then
$$
\frac{d V^{1/3}}{dt} =
 \frac{1}{12}
  (36 \pi)^{1/3}
  \sum_i v_i v_{i\ t}
   [ n_i (t=0) - N \xi_i(0) \frac{V}{\bar{v}(\vec{\xi}(0))} - n_{i\ \infty}
   \xi_i(0)
f_i(\vec{\xi}(0)) ]
$$
This equation can be easily integrated, which gives
$$
\int_0^{V^{1/3}}
\frac{dV^{1/3}}{ \frac{1}{12}
  (36 \pi)^{1/3}
  \sum_i v_i v_{i\ t}
   [ n_i (t=0) - N \xi_i(0) \frac{V}{\bar{v}(\vec{\xi}(0))} - n_{i\ \infty}
   \xi_i(0)
f_i(\vec{\xi}(0)) ]}
= t
$$

{ \bf Approximation of the characteristic concentration }

Consider (\ref{rr}) and put in  this equation all $\xi_i$ to their
characteristic values $\xi_{i(ch)}$ during the most intensive
consumption of the metastable phase. The characteristic value of
concentration has to be chosen as some artificial parameter.
sometimes it is useful to choose the mean value between the
initial and the final concentrations (the final concentration can
be found without explicit solution of the evolution equation; it
will be done later). Then
$$
\frac{d V^{1/3}}{dt} =
 \frac{1}{12}
  (36 \pi)^{1/3}
  \sum_i v_i v_{i\ t}
   [ n_i (t=0) -
N \xi_{i(ch)} \frac{V}{\bar{v}(\vec{\xi}_{(ch)})} - n_{i\ \infty}
   \xi_{i(ch)}
f_i(\vec{\xi}_{(ch)}) ]
$$
This equation can be easily integrated
$$
\int_0^{V^{1/3}}
\frac{dV^{1/3}}{ \frac{1}{12}
  (36 \pi)^{1/3}
  \sum_i v_i v_{i\ t}
   [ n_i (t=0) - N \xi_{i\ (ch)}
\frac{V}{\bar{v}(\vec{\xi}_{(ch)})} - n_{i\ \infty}
   \xi_{i\ (ch)}
f_i(\vec{\xi}_{(ch)}) ]} = t
$$

{\bf Approximation of stationary concentration}

Equation (\ref{xxi}) is a typical relaxation equation, which leads
to the relaxation of $\xi_i$ to the "bottom" of kinetic channel.
So, it is reasonable to put $\xi_i$ to the stationary value
$\xi_{i\ k}$ determined by
\begin{equation}\label{xxi1}
0
=
[ (n_i(0) - N \xi_i \frac{V}{\bar{v}} -
n_{i\ \infty} \xi_i f_i ) v_{t\ i} -
\xi_i \sum_j
(n_j(0) - N \xi_j \frac{V}{\bar{v}} -
n_{j\ \infty} \xi_j f_j ) v_{t\ j}
]
\end{equation}
The stationary value depends on the value of $V$, i.e. $\xi_{i\ k}
=\xi_{i\ k} (V)$, because $V$ characterizes the power of the
mother phase consumption by droplets.

Then equation for evolution of $V$  will be the following
\begin{eqnarray}
\frac{d V^{1/3}}{dt} =
 \frac{1}{12}
  (36 \pi)^{1/3}
  \sum_i v_i v_{i\ t}
   [ n_i (t=0) - N \xi_{i\ k} (V)
    \frac{V}{\bar{v}(\vec{\xi_k}(V))} -
\nonumber
\\
\nonumber
n_{i\ \infty}
   \xi_{i\ k}(V)
f_i(\vec{\xi_k}(V)) ]
\end{eqnarray}
This equation can be easily integrated,
which gives
$$
\int_0^{V^{1/3}}
\frac{dV^{1/3}}{ \frac{1}{12}
  (36 \pi)^{1/3}
  \sum_i v_i v_{i\ t}
   [ n_i (t=0) - N \xi_{i\ k}(V) \frac{V}{\bar{v}(\vec{\xi_k}(V))}
    - n_{i\ \infty}
   \xi_{i\ k}(V)
f_i(\vec{\xi_k}(V)) ]}
= t
$$

This solution is a very good approximation to the real  solution.
In all situations when there is no other big or small parameters
(rather often the ratio $n_{i \ \infty} v_{t\ i} /n_{j \ \infty}
v_{t\ j}$ is a small parameter\footnote{This will be considered
later.}, where $i$ and $j$ are two different components) this
approximation works well.

Certainly the system of algebraic equations (\ref{xxi1})
together with $\sum_i \xi_i = 1$ has to be
solved, which isn't  a simple task.

Generally speaking one can see that
the problem is to get the true value for concentrations. Then the
problem becomes much more simple.
One can formulate here the following statement:

\begin{itemize}
\item (F) When some approximation $\xi_{i \ appr}$ for the
concentration is adopted  then the kinetic equations can be
integrated as
$$
\int_0^{V^{1/3}}
\frac{dV^{1/3}}{ \frac{1}{12}
  (36 \pi)^{1/3}
  \sum_i v_i v_{i\ t}
   r_i}
= t
$$
where
$$
r_i=[ n_i (t=0) - N \xi_{i\ appr}(V) \frac{V}{\bar{v}(\vec{\xi_{appr}}(V))}
    - n_{i\ \infty}
   \xi_{i\ appr}(V)
f_i(\vec{\xi_{appr}}(V)) ]
$$

\end{itemize}

Now we shall analyze approximations, which are suitable at
different
periods of condensation. One can
define initial period of condensation and
the final period of condensation.
The initial period is the period, when the
quantity of the substance in the droplets is small in
comparison with
the total quantity of surplus substance.  Initial period covers
nucleation, which is appearance of droplets,
but does not coincide with nucleation.

The final period is the period when the quantity of remaining
surplus substance
in vapor is
small in comparison with the quantity of substance in droplets.

These definitions will be slightly modified later.

The  values which correspond to the transition\footnote{The end of
this transition is imaginary.} of all\footnote{The
density
of vapor
equals to the density of vapor saturated over the plane surface.}
surplus substance\footnote{May be the concentration is fixed.}
into the droplets will be called the final values and will be marked
by the lower index $fin$.

Between the initial and the final periods the intermediate period
takes place. Namely during this
period the macroscopic manifestation of a
phase transition takes place.

{\bf Approximation of different stages}

One can analyze the characteristic features of the last solution.
During
some period (initial period) one can see that $V$ grows rather
fast but $V$ is relatively small and in $r_i$ one can neglect $V$.
Then
$$
\int_0^{V^{1/3}}
\frac{dV^{1/3}}{ \frac{1}{12}
  (36 \pi)^{1/3}
  \sum_i v_i v_{i\ t}
   \hat{r}_i}
= t
$$
where
$$
\hat{r}_i=[ n_i (t=0)
    - n_{i\ \infty}
   \xi_{i\ k}(V=0)
f_i(\vec{\xi_{k}}(V=0)) ]
$$
Then
$$
\frac{V^{1/3}}{ \frac{1}{12}
  (36 \pi)^{1/3}
  \sum_i v_i v_{i\ t}
   \hat{r}_i}
= t
$$
The values of $\xi_{i\ k}$ are known - they are the coordinates of
kinetic channel at initial values of $n_i$, i.e. $n_i (t=0) $. So, there
are no problems in integration of the last equation. The
denominator doesn't depend on $V$. As the result  $V \sim t^3$.
The next step in approximation is the following:
The value of $V$ grows rather rapidly. Then in $r_i$ there appears a
rapidly
growing
 function $V$, which turns $ \sum_i v_i v_{i\ t}
   r_i$ to zero
$$
 0=  \sum_i v_i v_{i\ t}
   r_i \ \ .
$$
   Then $V$ stops to grow  and in the system one can see  some
   relaxation processes, which can be described as the final
   period. This period is characterized by some approximately
   constant and, thus, prescribed value of $V
   \equiv V_{fin}$.

At the final period
the value of $V$ becomes the function of $\vec{\xi}$ and
can be found from the last equation,
   which can be written as
$$
0=  \sum_i v_i v_{i\ t}
[ n_i (t=0) - N \xi_{i}(V) \frac{V}{\bar{v}(\vec{\xi}(V))}
    - n_{i\ \infty}
   \xi_{i}(V)
f_i(\vec{\xi}(V)) ]
$$
Later we shall write that $V= V_{fin} (\vec{\xi})$.

Now equations on $\vec{\xi}$ can be written as
\begin{eqnarray}\label{xxi2}
\frac{d\xi_i}{dt}
=
\frac{\bar{v}(\vec{\xi})}{V^{1/3}_{fin} (\vec{\xi})}
\frac{1}{4} (36 \pi)^{1/3}
[ (n_i(0) - N \xi_i \frac{V_{fin} (\vec{\xi})}{\bar{v}} -
\nonumber
\\
\\
\nonumber
n_{i\ \infty} \xi_i f_i ) v_{t\ i} -
\xi_i \sum_j
(n_j(0) - N \xi_j \frac{V_{fin} (\vec{\xi})}{\bar{v}} -
n_{j\ \infty} \xi_j f_j ) v_{t\ j}
]
\end{eqnarray}
This equation is the ordinary relaxation equation but certainly it
is nonlinear  and multi-dimensional. The analytical solution in
the general case is unknown but the relaxation to the stationary
value, which satisfies  the system of algebraic equations
\begin{eqnarray}\label{pop}
[ (n_i(0) - N \xi_i \frac{V_{fin} (\vec{\xi})}{\bar{v}} -n_{i\ \infty} \xi_i f_i ) v_{t\ i}
\nonumber
\\
\\
\nonumber
 -
\xi_i \sum_j
(n_j(0) - N \xi_j \frac{V_{fin} (\vec{\xi})}{\bar{v}} -
n_{j\ \infty} \xi_j f_j ) v_{t\ j}
]
= 0
\end{eqnarray}
is rather obvious.

In the binary case there are no problems in solution of kinetic
equation (\ref{xxi2}). It can be rewritten as
equation on one concentration $\xi_1$
$$
\frac{d\xi_1}{dt} =
\frac{\bar{v}^{2/3}}{\vartheta_{fin}^{1/3}}
\frac{(36\pi)^{1/3}}{4} R(\xi_1)
$$
where
\begin{eqnarray}
R(\xi_1) \equiv
(1-\xi_1) [(n_1(0) - N \xi_1 \vartheta_{fin} - n_{1\ \infty}
\xi_1 f_1(\xi_1)) v_{t\ 1} -
\nonumber
\\
\nonumber
\xi_1
(n_2(0) - N (1-\xi_1) \vartheta_{fin} - n_{2\ \infty}
(1-\xi_1) f_2(1-\xi_1)) v_{t\ 2}
]
\end{eqnarray}
and in approximation $\vartheta_{fin} = const$ this
equation can be easily integrated
$$
\int \frac{d\xi_1}{\frac{(v_1 \xi_1 +
v_2 (1-\xi_1))}{V_{fin}^{1/3}}
\frac{(36\pi)^{1/3}}{4} R(\xi_1)} = t
$$

Here appears
a problem what extensive variable we have to fix in the last
relation. When we come from $V_{fin}$ to $\vartheta_{fin}$ we have
$$
\int \frac{d\xi_1}{
\frac{(v_1 \xi_1 +
v_2 (1-\xi_1))^{2/3}}{\vartheta_{fin}^{1/3}}
\frac{(36\pi)^{1/3}}{4}
R(\xi_1)
}
= t
$$
It seems reasonable to choose that extensive variable, which variations are
small. The problem is what coefficients $\lambda_i$
in extensive variable $\sum_i \lambda_i \nu_i$ we have to choose
(for $\vartheta$ all $\lambda_i = 1$, for $V$ we
have $\lambda_i = v_i$). Certainly, we have to choose\footnote{With restriction
 $\sum_i \lambda_i $ = 1.} $\lambda_i$ to have the
smallest
variations\footnote{It can be derived in
some initial approximation.} of  $\sum_i \lambda_i \nu_i$.
Here we can
repeat
all previous constructions. For the reason of physical clearness
we shall conserve below $V$ (and $\vartheta$), but it is more
accurate to take $\sum_i \hat{\lambda}_i \nu_i$.
In any case this problem is rather artificial because the
concentrations approach to the stationary values.

The possibility to get analytical solution in the binary case
allows to suggest the pair approximation. We choose the fastest
component (due to the kinetic coefficient). All other components
will be the second component\footnote{Or one can divide all components into two groups.}.
 Then one can get the solution
presented above. The concentration of the given fast component
will approach to the final value. Then the same can be done with
the set of the rest (formally slow) components. This loop can be
repeated up to the end.

During this relaxation such an interesting phenomena
as the partial disappearance of some components from
droplets can take place.
Really, suppose that there is a fast component and a slow
component. The chemical potential of a fast component has a
minimum near $\vec{\xi_{k}}(V=0)$. The chemical potential of the
slow component has a giant minimum far from $\vec{\xi_{k}}(V=0)$.
What will happen in the system?
At first $\vec{\xi} \approx \vec{\xi_{k}}(V=0)$ and the density of
vapor of the first component approaches to the value corresponding
to the minimal value of the chemical
potential for the fast component. But later the slow
component saturates the droplets and this rises the chemical
potential of the fast component from the minimum. Then the chemical
potential of the fast component in the vapor has to grow. It means
that the density has to grow and there is a partial disappearance
of the fast component from droplets back to the vapor phase.

One can speak about {\bf the reverse phase of the phase transition
for some components.} Certainly, this is only the first peak of
the possible oscillatory regime, which can take place here. This
example shows us that many interesting phenomena were thrown out
in a wrong solution presented in \cite{djikaiev2}. This
redistibution of substances in droplets has nothing in common with
redistribution of substance in droplets  during the coalescense described
by Lifshits and Slezov \cite{Lifshic-Slezov}.

As the result one can speak about the inertial phenomena in the
phase transition. This conclusion  takes place in the
multi-dimensional decay kinetics and is
the principally new feature of
the phase transition.

We have extracted the initial period of condensation and the final
period of condensation.
We recall that the period between
the initial one and the final one will be called the
the intermediate period\footnote{For every component the positions of
these periods are specific.}.
 Do the initial and the final periods cover practically
the whole period of
condensation before the coalescense? Due to the avalanche
consumption of surplus substance  (or due to the rapid growth of
$V$ at the initial stage) one can establish the following very
important fact
\begin{itemize}
\item The intermediate period between the initial period and the
final period is rather short (in comparison with the initial
period).
\end{itemize}

Then with a small relative error one can  speak
about the boundary between the initial period and the final period
and require the coincidence of corresponding solutions at this
boundary\footnote{To get this boundary one has to prolong the
solution at the initial period to the intermediate period.
If we prolong the asymptotic solution we can  incorrectly get
the long
intermediate period.}. This gives the values of arbitrary constants in the
solution at the final period.

Now we shall discuss these periods in details.

{\bf Initial stage}

To give
the detailed  description of the initial stage we shall
write kinetic equations. It can be done as
\begin{equation}
\frac{d\nu_i}{dt} =
\frac{1}{4}
v_{i\ t}  (36 \pi)^{1/3} (\sum_j v_i \nu_i)^{2/3} [ n_i (t=0)   - n_{i\ \infty} \xi_i
f_i(\vec{\xi}) ]
\end{equation}
We see that here
$$
\xi_i = \xi_{i\ k}(t=0)
$$
Then
$$
\nu_i^{1/3}  = \Lambda_i t
$$
The same has been done in investigation of the nucleation period,
but it is clear that this description can be prolonged on the
whole initial stage (normally, nucleation period belongs to initial stage).
One has to note that in \cite{djikaiev2} there is no clear picture
of the characteristic periods and stages during the condensation.
Moreover there are many errors. For example it is stated that at
the first period (nucleation period) the densities of components
in vapor were considered to be constant (\cite{djikaiev2},
page 391). This
conclusion is wrong. The densities are functions of
time, the simplification is approximate equality of the densities
to some constant values. But in consideration of nucleation, which
is very sensible to the values of densities one can not
consider these densities as
to the constant values. Here the requirement of precision is more
weak\footnote{ Except the situation with the secondary nucleation,
which will be considered later.}
  - we need only the high relative accuracy of the densities defining. So,
here we can put them to constants and consider the initial stage.

One can use another possibility to enlarge the region of
description by the analytical solution. This possibility lies in
the approximation of high supersaturations.
Let us consider this approximation.

In one component condensation one can show that the  intensive
condensation is possible only when
$$
n/n_{\infty} \gg 1
$$
This conclusion  comes from the standard Gibbs rule
which states that  the free energy of the critical
embryo is three times less than the surface energy of the same
embryo.
 If
$n/n_{\infty} \rightarrow 1 $   then  $F_c$ is too big and the
nucleation rate, which is proportional to $\exp(-F_c)$ is
negligibly small\footnote{The
value of surface tension referred to
one molecule can not be too small otherwise
one has to take into account the fluctuations of the surface and
the situation will be close to the second order phase transition.
So, here there is a hidden contradiction between the intensive
nucleation and the possibility of thermodynamic
description of the critical embryo.}.
Ordinary $n/n_{\infty} \approx 3 \div 5 $.

In the situation of multicomponent condensation this requirement
($n/n_{\infty} \gg 1$) isn't necessary because there are many
components. But still there can exist some leading components, for
which $n_i/n_{i\ \infty} \gg 1$. Then one can neglect $ n_{i\
\infty} \xi_i f_i(\vec{\xi})$ in comparison with $n_i - N \nu_i$
and then get
$$
\frac{d\nu_i}{dt} =
\frac{1}{4}
v_{i\ t}  S [ n_i (t=0) - N \nu_i ]
$$
Analogously one can get a more simple equation on $\xi_i$
\begin{equation}\label{xxi5}
\frac{d\xi_i}{dt}
=
\frac{\bar{v}(\vec{\xi})}{V^{1/3}}
\frac{1}{4} (36 \pi)^{1/3}
[ (n_i(0) - N \xi_i \frac{V}{\bar{v}} ) v_{t\ i} -
\xi_i \sum_j
(n_j(0) - N \xi_j \frac{V}{\bar{v}}
 ) v_{t\ j}
]
\end{equation}
Then the equation on the coordinates of kinetic channel will be
$$
[ (n_i(0) - N \xi_i \frac{V}{\bar{v}} ) v_{t\ i} -
\xi_i \sum_j
(n_j(0) - N \xi_j \frac{V}{\bar{v}}
 ) v_{t\ j}
]
=
0
$$
Then
the
value
$$
(n_i(0) - N \xi_i \frac{V}{\bar{v}} ) v_{t\ i}/\xi_i \equiv c
$$
is some invariant
(it doesn't depend on component).

It is clear that the channel will be purely kinetic. It is
easy to see the concentrations $\xi_i$ can not approach to $0$,
because then the equality
$$
n_i(0)  = \xi_i ( c /v_{t\ i} + N  \frac{V}{\bar{v}} )
$$
will be violated\footnote{When $\bar{v}$ is the regular function
of concentrations.}.

One can easily get equation on $c$.
Really,
$$
\xi_i = \frac{n_i(0)}{c /v_{t\ i} + N  \frac{V}{\bar{v}}}
$$
Then one can get equation for $\bar{v}$ via $c$
$$
\bar{v} = \sum_i v_i \frac{n_i(0)}{c /v_{t\ i} + N  \frac{V}{\bar{v}}}
$$
Then $\bar{v} = \bar{v}(c)$ and
because of $\sum_i \xi_i = 1$ one can get equation on $c$
$$
1 = \sum_i  \frac{n_i(0)}{c /v_{t\ i} + N  \frac{V}{\bar{v}(c)}}
$$
This equation can be easily solved. Ordinary it has a unique
solution.

As the result we see that to observe the kinetic channel near
$\xi_i = 0$ it is necessary to have really giant  values of
$f_i$. It means in this case that the values of $\xi_i$ are
governed by $f_i$ and they are rather insensitive to the vapor
depletion. Then $\xi_{i\ k}$ is the quasiintegral of evolution and the
stationary approximation for $\xi_{i\ k}$ suggested earlier is  good.
 Then it is easy to see the
relaxation to this value.

{\bf Final stage}

The final stage can be described by approaches described earlier.
Here we shall see that the end of this stage can be described
extremely simple. The final values (they can not be attained
because of coelescense) can be gotten as solutions of the
algebraic system
$$
d\nu_i
/ dt = 0
$$
or
\begin{equation} \label{final}
0 =
 n_i (t=0) - N \nu_i - n_{i\ \infty} \xi_i
f_i(\vec{\xi})
\end{equation}
They will be marked as $\nu_{i\ fin}$.
When $\nu_i$ are close to $\nu_{i\ fin}$, then
the kinetic equations
for
$\Delta \nu_i = \nu_i - \nu_{i\ fin}$
can be rewritten as
$$
\frac{d \Delta \nu_i } { dt} =
- \frac{1}{4} v_{t\ i} N S_{fin} \Delta \nu_{i}
$$
where
$$
S_{fin} = (36 \pi)^{1/3} (\sum_j v_{j} \nu_{j\ fin}^{(k)})^{2/3}
$$
These equations  are linear and can be easily solved.
As the result we have the solution at the end of the
final stage.
The solution is
$$
\Delta \nu_i  = c_i  \exp(- \frac{v_{t\ i}S_{fin} N  t}{4})
$$
with generally unknown  negative constants $c_i$. They can be determined by
coincidence of solutions at initial stage and at final stage at
the boundary between these stages.

This completes the general   investigation of the
one-stage nucleation.

In the next section when we shall study the multi-stage nucleation
process. We shall give another approximate solutions, which will be
valid under some specific situations of small concentrations.

\section{Nucleation in the case of small concentration in
droplets}

Why shall we pay a special attention to the further investigation
of nucleation? One can recall one important feature of one
component nucleation. The one component theory predicts that the
number of droplets $N$ is proportional to the amplitude of
spectrum $f$ in the power $3/4$. Since $f$ is proportional to
$\exp(-\Delta F_c)$ it means that for two channels with $\Delta_1
F_c < \Delta_2 F_c - 1$ the strong inequalities $f_1 \gg f_2$ and
$N_1 \gg N_2$ take place. This conclusion in the multicomponent
condensation isn't valid because one has to take into account
specific kinetic factors which can be different for different
channels.

One can also notice another important feature. In the
multicomponent theory the positions of thermodynamic channels
don't observe the order. When
in the one component case we analyze the dependence
$F(\vartheta)$ along the thermodynamic channel we see that the
higher is activation barrier the higher lies the curve
$F(\vartheta)$ (along the bottom of thermodynamic channel). But
since the values of surface tensions in different thermodynamic
channels are different one can come to situation when the channel
with the highest barrier (i.e. wiht the highest $F_c$) later
becomes the lowest channel\footnote{We compare channels $\vartheta$ being fixed.}.
It means that we have
to analyze all channels despite the heights of their barriers.

At first we shall describe the ordinary situation.
The process of nucleation is ordinary localized in time. Really,
already in the initial period
$$
\nu_i = \Lambda_i^3 t^3
$$
and
$$
n_i (t) = n_i (t=0) - N \Lambda_i^3 t^3 \ \ .
$$
Thus, every component leads to the fall of intensity
of the droplets formation in the channel number  $j$ (let us
enumerate the channels according the heights of activation barriers)
as
$$
J(t) = J(t=0) \exp(
\sum_i \nu_{c\ i\ j} ( n(t) -n_i(t=0) )) =
 J(t=0) \exp(-
\sum_i \nu_{c\ i\ j} N \Lambda_i^3 t^3 ) \ \ ,
 $$
 where $\nu_{c\ i\ j}$ is the number of molecules of $i$ component
 in the critical embryo for the  channel number $j$.
 This rapid decrease of the nucleation rate $J$ according to the
 previous relation means that there will be no further formation
 of droplets. So, the process of intensive nucleation is finished in all channels.
 This situation has  been already described in the first part of the previous
 section.

But still there is  one important exclusion from this situation.
We shall consider the nucleation in some channel which will be
called as the second one. So, the last index now will merk the
channel of nucleation.
When the concentration $\xi_i $ of $i$-th component in the
droplets in the first channel is small,
 then $\Lambda_i$ can be small
and at moderate times $t$ the value $-\nu_{c\ i\ 2} ( n(t)
-n_i(t=0) )$ is small enough when $\nu_{c\ i\ 2}$ isn't too big\footnote{We suppose this
property for the second channel.}.
 So, the intensive nucleation (in
comparison with initial value) in the second channel continues
even when the intensive nucleation in the first channel stops.
Then the number of droplets appeared in the second channel becomes
important and it is interesting to calculate this number.

The reason of termination of
intensive nucleation in the second channel can be the vapor
depletion by both droplets appeared in the first channel and droplets
appeared in the second channel. This situation becomes complex and
requires a special investigation. It will be given in this
section.

From the first point of view it seems that this situation is
rather artificial. But as has been already noticed in introduction
when we have two immiscible   substances we come to two separate
processes of nucleation - the first process is the nucleation of
the
 first substance into the droplets of the first substance
and the second process is the nucleation of the second substance
into the droplets of the second substance. These processes have
nothing in common. This situation has to be described in frames
of the binary condensation also and  namely this situation
falls into the case, which will be considered.
When there is a small miscibility of substances we have a
moderate influence of the first process on the second one and have
to take this interaction into account.

We shall call the situation when several channels of nucleation
are sequentially important as the "long nucleation".

Now we shall investigate the case of long nucleation in details.
We shall describe here the two component case, then the necessary
remarks for generalization will be given.

We have already noticed the possibility of partial reverse
condensation and shall call this as "inertial" effects.

At first  we shall analyze the general features of the process,
then we shall give the theory for the process with no "inertial"
effects in the first channel, then the situation with "inertial"
effects in the first channel will be described.

\subsection{General features of the "long nucleation"}

The nucleation process in the first channel has been already
described and  we know the number $N_1$ of the droplets
appeared during the nucleation in this channel (earlier this
quantity was denoted by $N$).
Let us suppose that in the first kinetic channel $\xi_{1\ k} $ is
small. In the second channel $\xi_{2\ c}$ is small.
One can justify the following statement:
\begin{itemize}
\item {\it Up to the end of nucleation in the second channel the
droplets appeared in the second channel can not essentially
perturb the rate of growth of droplets appeared in the first
channel. }

Then we can investigate the separate process of vapor
depletion by the droplets appeared in the first channel.

\end{itemize}

For the evolution of droplets appeared through the first channel
 $\xi_1 \ll 1$, $\xi_2 =1$, the values of concentrations are
known.

The system of kinetic equations for this problem can be written in
approximation of ideal solution since the solution is really the
dilute one. The system of equations  is given by\footnote{For
dilute solution one can use the approximation of the ideal
solution (may be with renormalized parameters).}
$$
\frac{d\nu_1}{dt} = \frac{v_{t\ 1}}{4} S  (n_1(0) - N \nu_1 -
n_{1\ \infty} \xi_1) \ \ ,
$$
$$
\frac{d\nu_2}{dt} = \frac{v_{t\ 2}}{4} S  (n_2(0) - N \nu_2 -
n_{2\ \infty} ) \ \ , $$
$$
S = (36 \pi)^{1/3} (v_1 \nu_1 + v_2 \nu_2)^{2/3} \ \ .
$$
The value of $S$  when there is no extremely small values in
ratios $v_i/v_j$ can be presented as
$$
S= (36 \pi)^{1/3} (v_2 \nu_2 )^{2/3} \ \ .
$$

Then it is possible to integrate these equations\footnote{In the first equation
one can neglect $n_{1\ \infty} \xi_1$, which leads to
$$
\frac{d\nu_1}{dt} = \frac{v_{t\ 1}}{4} S  (n_1(0) - N \nu_1 )
$$
and the first equation can be then integrated in a very simple way.} . Equation for
$\nu_2$ will be
$$
\frac{d\nu_2}{dt} = \frac{v_{t\ 2}}{4} (36 \pi)^{1/3} (v_2 \nu_2
)^{2/3}  (n_2(0) - N \nu_2 - n_{2\ \infty} ) \ \ .
$$
Since the r.h.s. doesn't contain $t$ this equation can be
integrated
$$
\int_0^{\nu_2} \frac{d \nu_2}{\frac{v_{t\ 2}}{4} (36 \pi)^{1/3}
(v_2 \nu_2 )^{2/3} (n_2(0) - N \nu_2 - n_{2\ \infty} ) } =  t\ \ .
$$
So, now $\nu_2$ is the known function of $t$.
Then equation for $\nu_1$ can be written as
$$
\frac{d\nu_1}{dt} =
\frac{v_{t\ 1}}{4}(36 \pi)^{1/3} (v_2 \nu_2(t) )^{2/3}
  (n_1(0) - N \nu_1 - n_{1\ \infty} \xi_1)
$$
and
since $\xi_1 \approx \nu_1/\nu_2$
it can be rewritten as
$$
\frac{d\nu_1}{dt} =
\frac{v_{t\ 1}}{4}(36 \pi)^{1/3} (v_2 \nu_2(t) )^{2/3}
  (n_1(0) - N \nu_1 - n_{1\ \infty} \nu_1 / \nu_2(t)) \ \ .
$$
Since it is the first order linear differential equation it can be
easily integrated by the known formulas.

This solution is explicit, but the model with the small
concentration and ideal solution can be criticized.
One can see that the limit of the last solution is
$$
n_1 (t) \rightarrow  0 \ \ .
$$
The final values of $\nu_{i\ fin}$ are obtained on the base of
algebraic equations (\ref{final}) and they doesn't depend on
$v_{t\ i}$ and on ratios\footnote{ The initial positions
of kinetic channels strongly depend on $v_{t\ i}$ and $n_{i\
\infty}$. } of $n_{i\ \infty}$. So, one can easily come to the
situation when $\xi_{1\ fin}$ doesn't go to $0$. The last
solution doesn't satisfy this condition.

The last remark means that we have to investigate the general
situation.
From the previous section we know that the solution can
be composed from the part corresponding to the initial period and
the part corresponding to the relaxation during the final period.
So, there appeared two characteristic problems:
\begin{itemize}
\item
1). To describe nucleation in the second channel during the initial
period
\item
2). To describe nucleation in the second channel during the
final period.
\end{itemize}

The last problem has to be solved in different situations.
When\footnote{The final values here and later are taken only
with account of  droplets appeared through the first channel (this
number is already known).}
$$
\xi_{1\ k} f_1(\xi_{1\ k}) >\xi_{1\ fin} f_1(\xi_{1\ fin})$$
then  ordinary there are no inertial effects\footnote{
Certainly, the opposite situation can take place.}.
The rhs in the last
inequality can be taken at initial values of vapor densities since
as it has been noticed the channel in the situation of small
concentration is very deep and concentration is here a quasiintegral of
evolution. One can see  here that the value $n_1$ constantly
decreases in time.
 Here one has to investigate the
nucleation in the second channel during
the relaxation with decrease of density.

When
$$
\xi_{1\ k} f_1(\xi_{1\ k}) < \xi_{1\ fin} f_1(\xi_{1\
fin})
$$
 one can come to the situation when $n_1$ decreases at
first in time to the value  $\xi_{1\ k} f_1(\xi_{1\ k}) $. The
value of $n_1$  can not come too close to this value.  Then it begins
to increase and approaches to $\xi_{1\ fin} f_1(\xi_{1\
fin})$.
So, beside the problem to describe the nucleation during the
decrease of $n_1$  it
is necessary to describe  the nucleation during the increase of
$n_1$. This problem also splits into two separate
problems  to describe the evolution when there is essential
increase of $n_1$  and to describe evolution when
$n_1$ already attains the asymptotic value.

It is extremely important to note that we need the solution only
when $\xi_1 \ll 1$. Only when $\xi_1 \ll 1$ we can suppose that
there will be rather intensive droplets formation in the second
channel.  At the same time only then the  solution will have
a very high precision,
which is necessary to describe the nucleation in the second
channel.

\subsection{Secondary nucleation during the initial stage}

We shall call the appearance of droplets through the second
channel as the secondary nucleation.
During the initial stage of the primary condensation the equation
for the density $n_1$ as function of $t$ will be
$$
n_1 (t) = n_1(0) - N_1 (\Lambda_1 t)^3 - g_{2\ 1} \ \ .
$$
Here $N_1$ is the total number of droplets appeared through the
first  channel (since the nucleation through the
first channel is over we know this number), $g_{2\ 1}$ is the
number of molecules of the first component in the droplets
appeared through the second  channel. Then having
repeated for the second channel the same procedure as we had done
for the first channel we get the following equation for
the number of the molecules of the first component $g_{2\ 1}$
in the droplets appeared through the second channel
$$
g_{2\ 1} (t) =
J_{2\ *}  \Lambda_{1\ 2}^3
\int_0^t dt' (t-t')^3 \exp(-  [\sum_i
\nu_{i \ 2 \ c}(\tau) N_1 \Lambda_i^3 t'^3
+
$$
$$
\sum_i \nu_{i \ 2 \ c} (\tau)
\xi_{i\ k} g_{2\ 1}(t')/ \xi_{1\ k}
] )
$$
where $\tau=0$ or $\tau$ is some characteristic value, may be
at the asymptotic solution.
Here $\Lambda_{1\ 2}$ is the  constant in the law of growth
$$
\frac{d\nu_1^{1/3}}{dt} = \Lambda_{1\ 2}
$$
of the molecules of the first component in the droplets appeared
through the second channel, $J_{2\ *}$ is the amplitude value of
the stationary rate of nucleation in the second channel, $\nu_{i \ 2\ c}$ is the number
of molecules of $i$-th component in the critical embryo in second
channel, $\Lambda_i$ is the constant in the law of growth
$d\nu_i^{1/3} / dt = \Lambda_i t $ of the molecules of the $i$-th
component in the droplets appeared through the first channel,
$\xi_{i\ k}$ is the concentration of $i$-th component molecules in
the second thermodynamic channel.

Having combined  the constants we have
$$
g_{2\ 1} (t) =
J_2 \int_0^t dt' (t-t')^3 \exp(-  A t'^3
- B g_{2\ 1}(t') )
$$
with known constants $J_2$, $A$, $B$. Both functions $\exp(-A
t'^3)$ and $\exp(-B g_{2\ 1}(t'))$ can be interpreted as step-like
functions, which allows to state that

{\it To get the rate of nucleation  the value of $g_{2\
1}$ in the r.h.s. can be calculated
approximately without account of $\exp(-A t'^3)$ in frames of the
first approximation in the standard iteration procedure}
$$
g_{2\ 1\ (0)} = 0
,\ \
g_{2\ 1\ (i+1)} (t) =
J_2 \int_0^t dt' (t-t')^3 \exp(-  A t'^3
- B g_{2\ 1\ (i)}(t') )
$$

So,
$g_{2\ 1\ (1)} =  J_2 t^4/ 4$ and then
$$
g_{2\ 1\ (2)} =
J_2  \int_0^t dt' (t-t')^3 \exp( - A t'^3 - B J_2  t'^4 /4)
$$
So, the rate of nucleation looks like
$$
\tilde{J} = J_2   \exp( - A t'^3 - B J_2 t'^4 /4)
$$
and it has the step like behavior. Moreover it is the product of
two step-like functions  $\exp(-A t'^3) $ and $\exp(-B J_2 t'^4/4)$.
So, we can introduce two characteristic sizes
$$
\Delta_a t = A^{-1/3}
$$
and
$$
\Delta_b t = (B J_2/4)^{-1/4}
$$
We have to take the minimal value from $\Delta_a t$ and $\Delta_b
t$ to determine the nucleation duration
$$
\Delta t = min \{ \Delta_a t , \Delta_b t \}
$$
Then the rate of nucleation as a function of time is
$$
J(t) = J_2 \Theta(t) \Theta(\Delta t - t)
$$
and the total number of droplets $N_2$ appeared through the second
channel is
$$
N_2 = \int_0^{\infty} J(t) dt = 0.9 J_2 \Delta t \approx J_2
\Delta t
$$
Now we have determined  all main characteristics
of phase transition.
These characteristics can be determined with the more high
accuracy using the perturbation technique and some iteration schemes.
Here it is important that the base for these constructions is
now given.

\section{The properties of further evolution and their effects
on the secondary nucleation}

Now we have to consider the situation when the secondary nucleation takes
place at times greater than the initial period. We know that at
some intermediate times after the end of initial period but before
the asymptotic final period the solution is known with essential
error. Fortunately one can prove the following statement, which
will help to give a simple analytical description:
\begin{itemize}
\item
{\it When the process of
the secondary nucleation stops after the end of the
primary nucleation
initial period but before the time, which two or three times
exceeds the time of the end of initial period of the primary nucleation
one can simply
neglect the vapor consumption by the droplets from the second
channel after the end of primary initial period. }

Here we define the initial and the final periods referred to the
corresponding components.
\end{itemize}

Now we shall explain this statement. At first we shall see that
the period between the initial period
and the asymptotic period   has a rather small (in
comparison with initial period) duration. We shall see this on
example of the one component  nucleation where the precise solution can be
easily obtained. Really, the evolution equation can be reduced
after the certain renormalizations to the following form
$$
\frac{dz}{dt} =
1 - z^3
$$
with the initial condition $z(t=0) = 0$. Here $z$ is the
renormalized coordinate of the
droplets linear size $\vartheta^{1/3}$.
The solution is the following
$$
\int_0^z \frac{dz}{1-z^3} = t
$$
and the integral can be taken analytically, which gives
$$
t=  - \frac{1}{6} \ln \frac{(z-1)^2}{z^2 + z+1}  +
\frac{1}{\sqrt{3}} \arctan
\frac{2z+1}{\sqrt{3}} -
\frac{1}{\sqrt{3}} \arctan
\frac{1}{\sqrt{3}}
$$
The last solution has the asymptotics
$$
t =  - \frac{1}{3} \ln\frac{1-z}{\sqrt{3} \exp(\sqrt{3} \pi /6)} \
\ \ \  z = 1- \exp(\sqrt{3}\pi/6) \sqrt{3} \exp(-3t)
$$
 at big $t$ and
\begin{equation}\label{smm}
 z = t- \frac{t^4}{4} + \frac{3}{4}
\frac{t^7}{7}
\end{equation}
at small $t$.

In Fig. 2 one can see the precise evolution and the asymptotics.
The curves are drawn for the value of excess of density, i.e. for
the normalized difference $\Psi \sim n - n_{\infty}$, which  is proportional
here to $1-z^3$. One can see one long curve, which binds the curves,
which are cut off. This long curve is the precise solution. The
short curves are asymptotics. There is one asymptotic $\hat{f}_{as}$ for
the big times and two asymptotics for the small times. The
asymptotics for the short times differ by the number of terms for
$z$ in (\ref{smm}), which are taken into account.  The curve
$\hat{f}_{(1)}$ corresponds to the account of the first two terms in
expression for $z$ in (\ref{smm}) and the curve $\hat{f}_{(2)}$
corresponds to the  account of the first three terms in
expression for $z$ in (\ref{smm}).

\begin{figure}



Figure 2. Different approximations to the evolution of density

\end{figure}

One can see that the period of time when both the long time
asymptotics and the short time asymptotics don't work has a rather
short duration. This duration has to be compared with the duration
of initial period. Since the precise solution is known one can
give the  estimates but everything is clear already
from the picture. One can only note that this property comes from
the big power $3$ in the law of droplets growth. This is the same
reason as for the property of the avalanche consumption
\cite{TMP}. So, one can here note the following important property
\begin{itemize}
\item
 {\it The period of consumption of the main surplus
substance is  rather short in comparison with the initial period.}
\end{itemize}

So, one can speak also about the avalanche consumption of the all
surplus substance.

This property is typical also for the case of the dynamic
conditions \cite{Novos}. Earlier it was only noted that the period
of consumption of all surplus substance is short in comparison
with the period of waiting (see \cite{Novos}). Now we can state
that the period of the avalanche consumption (or the intermediate period)
is short in
comparison with the period when the vapor density approximately
equals to the density at the moment of the most intensive
formation of droplets (or simply to the density at the moment of
intensive formation of droplets), i.e. to the initial period.

The small parameter characteristic for the initial period in the
solution of the last equation is $g=z^3$ in comparison with $1$.
One can see that when $m \gg 1$ the value $z^m$ will be small even
when $z$ is comparable with $z_{fin} = 1$ (for example $(1/2) =
1/8 \ll 1$ can be considered as the small parameter). The small
parameter for the asymptotic decompositions is\footnote{This
estimate is very rough, it depends on activity coefficients $f_i$
and can be established only in the order of magnitude.} $$
\frac{|z-z_{fin}|}{z_{fin}} \ll 1
$$
where $z_{fin}$ corresponds to $V_{fin}$.
 Here $z_{fin} = 1$ and it
means that simply $z$ is close to $1$. So, we see that there is
practically no gap between the initial and the final asymptotic
periods.

One has to outline that there is no satisfactory approximation for
the value of the vapor density or for the value of the number of
molecules in the liquid phase as a function of time because namely
during the intermediate period occurs the phase transition - the
main quantity of substance goes from the vapor to the liquid
phase.
Fortunately, these values aren't directly involved in the
secondary nucleation description.
But for the times of duration of characteristic periods,
which are necessary for the investigation of the secondary
condensation the required approximations are
rather precise.

In the multicomponent case the convincing arguments are the same.

Now we shall explain why it is possible to miss already the
two or three times of duration of initial period. Certainly the
density attained at the asymptotics $n_{i\ as}$ is essentially
smaller than the initial density. So, the rate of nucleation will
be many  times smaller than the initial  rate of nucleation.
The value of the critical number of molecules of given component
in the second channel
can not be very big (since the derivative of the free energy
barrier on density is this value) otherwise the rate of nucleation
at $n_{i\ as}$ will be absolutely negligible in comparison with
the initial rate of nucleation. Then the period of initial peak of
nucleation through the second channel will be not negligibly
 short
in comparison with the duration of initial stage. Then we see
that the number of droplets though the second channel appeared at
the asymptotic can be comparable with the number of droplets
through the second channel appeared in the peak of nucleation only
in one case - when the duration of the whole nucleation period
(including the nucleation at
the asymptotics) strongly exceeds the duration of the initial
period.

A special question which has been dropped out of  consideration
is the multiplicity of the final states.
It was supposed that the "final" is unique. But the algebraic
equations appeared from the conditions of the absence of the regular
growth  can have  several roots. It means
the several possible final states\footnote{"Final" states aren't the
real final ones, the process of coalescence will follow
the process of the regular growth of monodisperse spectrum.}.
This complicates the theoretical description. The first problem to
solve here is to see what final state corresponds to the given
channel of nucleation.  We have to construct  asymptotic for every
final state and then to decide what asymptotic is the closest one
to the solution at the initial period. this problem is the pure
algebraic one and can be solved without difficulties.

The second problem is the transition between final states.
Since the regular growth is absent here, we can return to the
free energies of the final states (or of the droplets in the
final states) $F_{fin}$. The rate of transition between final
states is proportional to $\exp(-\Delta F_{fin\ i\ j})$ where
$\Delta F_{fin\ i\ j}$ is the height of
transition barrier
(it depends on the numbers $i$ and $j$ of final states).
The
values
$\Delta
F_{fin\ i\ j}$ are so giant that ordinary there is no sense
to
consider transitions between final states.
But  in some peculiar situations when the height of transition
barrier is't too big, one can observe
a very interesting kinetic behavior, which forms the problem
for a separate publication.
This transition may
influence also on kinetics of coalescence\footnote{Certainly, the
coalescence of ground final state (with minimal $F_{fin}$) will be
described by the classical Lifshic-Slezov asymptotic.}, which
is also a very interesting point of investigation.
One can only note that transition between final states situated
on opposite
sides
of metastability gap has to go through the total evaporation
of embryos. This is very slow process, its characteristic times
strongly exceeds even the characteristic coalescence times.
Generally speaking, kinetics of evolution of final states
resembles in mathematical
structure of equations the micellar evoltuion and, thus, we need
not to give here the special description.

\subsection{Secondary nucleation during the relaxation period}

We need now  to investigate the evolution at big times,
corresponding to the asymptotic solution.

Here we have to take an explicit account of the initial peak of
metastability by introduction of total number of droplets appeared
through the second channel during this peak. This number is given
by
$$
 N_* = \int_0^{\infty} J_2 \exp(-A t'^3) dt' = 0.9 J_2
A^{-1/3}
 \ \ , $$
where this approximation and all parameters are described in the
previous sections.

We can here also introduce the duration of nucleation peak $\Delta
t$ as $$\Delta t = A^{-1/3}
\equiv \Delta_a t  \ \ . $$

At these big times one can write for $g_{2 \ 1}$ the following
expression
$$
 g_{2\ 1} = N_* z^3 +\tilde{f}_2 (z - \delta z)^4/4 \ \ .
$$
Here
$
\tilde{f}_2  =  J_{2\ s}
(\{ n_{i\ fin}  \} ) / \Lambda_{1\ 2} (\{ n_{i\ fin}\}) $
 is the stationary rate of nucleation $ J_{2\ s}
(\{ n_{i\ fin}  \} )$ through the second channel
 attained at the asymptotic
values (it can be calculated by known formulas for the stationary
rate of nucleation at known final asymptotic values) divided by
the rate of droplets
growth $\Lambda_{1\ 2} (\{ n_{i\ fin}\})$,
the value $\delta z$ is the shortage due to
the fact that initial period is already accounted (this value can
be also explicitly calculated, it is coordinate  $\nu_1^{1/3}$
attained by the droplets born at $t=0$ up to the end of initial
stage), $z$ is the coordinate $\nu_{1\ 2}^{1/3}$ attained by the
droplets born in the second channel at $t=0$ up to the current
time $t$, i.e. $z=\int_0^t \Lambda_{1\ 2} (\{ n_{i} (t) \}) dt
\rightarrow \Lambda_{1\ 2} (\{n_{i\ fin}  \}) t
$.

Since at asymptotic the concentration is constant and the
densities are also constant, then  the values $\nu_{i\ 2}^{1/3}$ grow in
time with constant velocity independent on $\nu_{i\ 2}$. Here
$\nu_{i\ 2}$ is the number of molecules of the $i$-th component in
the droplets born in the second channel. This explains the powers
$4$ in the previous expressions. Sometimes it is worth to add the
shift $\delta'z$ to the coordinate $z$ in the term $N_* z^3$ due
to the irregularity of droplets velocity at the initial stage and
at the transition from initial stage to the asymptotic solution
(when z is defined as $\Lambda_{1\ 2} (\{n_{i\ fin}  \}) t$).

Since $z= \Lambda_{1\ 2} (\{n_{i\ fin}  \}) t$ and $\delta'
z=\int_0^t ( \Lambda_{1\ 2} (\{n_{i} (t) \}) - \Lambda_{1\ 2}
(\{n_{i\ fin}  \} )) dt \rightarrow \delta' z=\int_0^{\infty}(
\Lambda_{1\ 2} (\{n_{i} (t) \}) - \Lambda_{1\ 2} (\{n_{i\ fin}  \}
)) dt $,
the
expression\footnote{The
value of $\delta z$ has to be reconsidered in the same manner.}
for $g_{2\ 1}$ will be the following $$ g_{2\ 1} =
N_* (z+\delta' z)^3 +
\tilde{f}_2 (z + \delta' z- \delta z)^4 /4  \ \ . $$

We see that the expression for $g_{2\ 1}$
is a very sharp function of
$z$. One can then speak here also about the avalanche consumption
of the metastable phase, which was outlined in \cite{TMP} as the
main feature of nucleation. This property allows to speak
about the upper boundary. To introduce the
definition of this boundary it is necessary to recall the form of
spectrum as function of time
$$
\tilde{f} =
\tilde{f}_2
\exp(-\sum_i \nu_{i\ c\ 2\
fin} \xi_{i\ fin} g_{2\ 1}/\xi_{1\ fin})\ \ .
$$
Again the avalanche character of vapor consumption can be proven.

Then the
boundary
$\Delta_f z$ is defined according to
$$
 \sum_i \nu_{i\ c\ 2\ fin} \xi_{i\
fin} g_{2\ 1}/\xi_{1\ fin} = 1
$$
 or
$$
 (\sum_i
\nu_{i\ fin} \xi_{i\ fin} )^{-1} =(N_* (z+\delta' z)^3
+\frac{f_2}{4} (z +\delta' z- \delta z)^4) /\xi_{1\ fin}\ \ .
$$
This
equation gives the value of $z$, which will be the boundary
$\Delta_f z$ of spectrum. The spectrum will be approximately
$$\tilde{ f
}= \tilde{f}_2 \Theta(z) \Theta(\Delta_f z -z) $$ and the number of
droplets is
$$
N_2 = N_* + 0.9   \tilde{f}_2 \Delta_f z
\approx N_* +    \tilde{f}_2 \Delta_f z  \ \ . $$

Ordinary the values  $\delta z$ and $\delta'z$ can be neglected
here.

This completes the
general investigation of the secondary nucleation but
below the more accurate methods will be presented.

\subsection{Description of the secondary nucleation at the
decreasing asymptotic}

During the asymptotic (final) period the concentrations $\xi_i$
 of the droplets
appeared in the first thermodynamic channel will approach the
final values according to the relaxation equation. Namely, the
final values $\xi_{i\ fin}$ satisfy the system of equations
\begin{eqnarray}\label{pop1}
[ (n_i(0) - N \xi_i \frac{V_{fin} (\vec{\xi_{fin}})}{\bar{v}}
-n_{i\ \infty} \xi_{i\ fin} f_i (\vec{\xi_{fin}}) ) v_{t\ i} -
\nonumber
\\
\\
\nonumber
 \xi_{i\ fin} \sum_j (n_j(0) - N \xi_{j\ fin}
\frac{V_{fin} (\vec{\xi_{fin}})}{\bar{v}} - n_{j\ \infty}
\xi_{j\ fin} f_j (\vec{\xi_{fin}})) v_{t\ j} ] \equiv 0
\end{eqnarray}
Then equation
(\ref{xxi2}) can be rewritten near $\xi_i = \xi_{i\ fin}$ as
\begin{equation}\label{xxi2l}
\frac{d\xi_i}{dt}
= \hat{L_i}(\vec{\xi_{ fin}}) (\xi_i - \xi_{i\ fin})
\end{equation}
where
$$
\hat{L_i} = \frac{\delta \hat{S}}{\delta \xi_i}
$$
and
\begin{eqnarray}
\hat{S} =
\frac{\bar{v}(\vec{\xi})}{V^{1/3}_{fin} (\vec{\xi})}
\frac{1}{4} (36 \pi)^{1/3}
[ (n_i(0) - N \xi_i \frac{V_{fin} (\vec{\xi})}{\bar{v}} -
\nonumber
\\
\\
\nonumber
n_{i\ \infty} \xi_i f_i ) v_{t\ i} -
\xi_i \sum_j
(n_j(0) - N \xi_j \frac{V_{fin} (\vec{\xi})}{\bar{v}} -
n_{j\ \infty} \xi_j f_j ) v_{t\ j}
]
\end{eqnarray}

Solution of the last system of equations is known
$$
\xi_i-\xi_{i\ fin} = C_i \exp(-\frac{t}{\delta_{i\ r} t}) \ \ ,
$$
$$
\delta_{i\ r}  t = |1/\hat{L}_i(\vec{\xi_{i\ fin}})|
$$
 with\footnote{If the relaxation really takes place, all $\hat{L}_i$ are negative.}
  constants $C_i$, which can be found from
the precise solution of (\ref{xxi}) by transformation to
asymptotics. These solutions lead to analogous exponential
solutions for $n_i - n_{i\ fin}$, namely $$ n_i-n_{i\ fin} =
C_{n\ i} \exp(-\frac{t}{\delta_{i\ n\ r} t}) $$
with constants $C_{n\ i}$ and corresponding relaxation times $\delta_{i\ r} t$.

Now we know rather precise asymptotics and can write the equation
for the nucleation rate in
the known exponential approximation
$$
J(t) = J_{2\ s} (\vec{n_{fin}}) \exp(\sum_i \nu_{i\ c\ fin}
[C_{n\ i} \exp(-\frac{t}{\delta_{i\ n\ r} t}) - \xi_{i\
fin}\frac{g_{2\ j}}{\xi_{j\ fin}}]) \ \ .
$$
Here the lower
index in $\nu_{i\ c\ fin}$ means that the critical value of
$\nu_{i}$ is taken at the final values of $n_i$. The value of
$g_{2\ i}$  is the number of molecules of the $i$-th
component in a liquid phase appeared through the second
thermodynamic channel.

At first we shall analyze the case of decreasing asymptotic when
 even without the
 natural decrease (due to $g_{2\ j}$) of the rate
of nucleation this value decreases in time.

Now we shall hold only the leading\footnote{The exponent with the greatest
relaxation time.} exponent, which brings the last
equation to
\begin{eqnarray}
J(t) = J_{2\ fin} \exp( \nu_{0\ c\ fin}
| C_{n\ 0} | \exp(-\frac{t}{\delta_{0\ n\ r} t}))
\nonumber
\\
\nonumber
 \exp( - \sum_i \nu_{i\ c\ fin}
 \xi_{i\ fin}\frac{g_{2\ j}}{\xi_{j\ fin}})
\end{eqnarray}
 $$
 J_{2\ fin} = J_{2\ s} (\vec{n_{fin}})  \ ,  $$
 where we have
marked the leading exponent by index $0$ (sometimes there are at
least two leading exponents with the same times of relaxation
since the sum of concentrations equals to one, so the amplitudes
have to be summarized). Then $$ J(t) = J_{2\ fin} \exp( \nu_{0\
c\ fin} | C_{n\ 0} | \exp(-\frac{t}{\delta_{0\ n\ r} t}))
 \exp( -
A_j g_{2\ j})
$$
where the constant $A_j$ is given by
$$
A_j =
 \sum_i \nu_{i\ c\ 2\ fin}
 \xi_{i\ fin}\frac{1}{\xi_{j\ fin}}
$$ Since ${g_{2\ j}}/{\xi_{j\ fin}}$ is invariant (it doesn't
depend on $j$), the rhs of expression
for $J$ doesn't depend on $j$ also. This
approximation can be used for further analysis.

At the asymptotics one can write $\nu_{i\ 2}^{1/3}
 = \Lambda_{i\ 2}
(t-t')$ where $t'$ is the time of formation of the droplet,
$$
\Lambda_{i\ 2} = (1/12)(36 \pi)^{1/3} \bar{v}^{2/3} \xi_{i\ k\
fin}^{-2/3} v_{i\ t} (n_{i\ as}  - n_{1\ \infty}
\xi_{i\ k\ fin} f_i
(\vec{\xi}_{k\ fin})) \ \ , $$
$\xi_{i \ k\ fin}$ is the
equilibrium concentration corresponding to the densities at the
asymptotic for the droplets appeared through the second channel,
$n_{i\ as}$ is the density at the asymptotic.

Now one can write equation for
$g_{2\ i}$ as
$$
 g_{2\ i} = N_* (z
+ \delta' z)^3
 +
 f_{2\ j} \int_{\delta z \sim 0}^z
dx (z-x)^3 \exp( \nu_{0\ c\ 2\  fin}
|C_{n\ 0}| \exp(-\frac{t'}{\delta_{0\ n\ r} t}))
 \exp( -
A_j g_{2\ j})
$$
where we specify that the nucleation is going through the second
channel.
Here $z= \Lambda_{i \ 2} t$, $x= z- \nu_i^{1/3}$, $t' =
x/\Lambda_{i\ 2}$, $f_{2\ i} = J_{2\ s}(n_{ fin})/\Lambda_{i\ 2}$.
Now this equation will be solved. It is easy to note that
$$
\int_{\delta z}^z dx (z-x)^3 \exp( \nu_{0\ c\ 2\ fin}
|C_{n\ 0}| \exp(-\frac{t'}{\delta_{0\ n\ r} t}))
 \exp( -
A_j
g_{2\ j})
$$
grows very rapidly (faster than $z^3$). Then the
spectrum of sizes is\footnote{Ordinary it is possible to neglect
$\delta z$, but even with $\delta z$ taken into account
the derivation will be quite similar. }
$$
 f  = f_{2\ j} \Theta(z) \Theta(\delta_0 z
- x) \exp( \nu_{0\ c\ 2\ fin}
|C_{n\ 0}| \exp(-\frac{t'}{\delta_{0\
n\ r} t}))
$$
plus the monodisperse peak from the initial period
 and $z(t) = \Lambda_{1\ 2} (t-t_0)$ where $t_0$ is
the time shift (it has to be associated
with $C_{n\ 0}$ analogously to $\delta z $, $\delta ' z$).
Here $\delta_0 z $ is the root of equation
$$
g_{j\ 2}(\delta_0 z)  = 1/A_j
$$
($j$ isn't important here).
 The total number of droplets is
approximated by
$$
N_2 = N_* + f_{2\ j}  \int_{\delta z}^{\delta_0 z }
dx
\exp(\nu_{0\ c\ 2\ fin} |C_{0 n}|
\exp(-\frac{x}{\Lambda_{i\ 2} \delta_{0\ n\ r} t }))
$$
 or after linearization
$$
N_2 =
N_* + (\delta_0 z - \delta z) f_{2\ j}
+ f_{2\ j} \nu_{0\ c\ 2\ fin} |C_{0\ n}| \delta_{0\ n\ r} t \Lambda_{i\ 2}
[ - \exp(-\frac{\delta_0 z}{\Lambda_{i\ 2} \delta_{0\ n\ r} t }) +
\exp(-\frac{\delta z }{\Lambda_{i\ 2} \delta_{0\ n\ r} t })]
$$

Here to solve equation on
$\delta_0 z$
 it is necessary to calculate $g_2(z)$ in appropriate  form.
One can done the following transformations: at first we shall
present $g_2$ as
$$
g_{2\ i} = N_* (z + \delta' z)^3
 +
 f_{2\ j} \int_{\delta z}^z dx (z-x)^3
(-1+\exp( \nu_{0\ c\ 2\ fin}
|C_{n\ 0}| \exp(-\frac{t'}{\delta_{0\ n\ r} t})
 \exp( -
A_j g_{2\ j}))
$$
$$
+ f_{2\ j} (z-\delta z)^4/4
$$
Then we can out the upper limit in the integral to $\infty$ and get
$$
g_{2\ i} = N_* (z + \delta' z )^3
 +
 f_{2\ i} \int_{\delta z}^{\infty}
dx (z-x)^3 (-1+\exp( \nu_{0\ c\ 2\ fin}
|C_{n\ 0}| \exp(-\frac{t'}{\delta_{0\ n\ r} t})
 \exp( -
A_j g_{2\ j}))
$$
$$
+ f_{2\ j} (z-\delta z)^4/4
$$
Then
$g_{2\ i}$
 is the polinomial of the forth power and we know the
analytical structure of $g_{2\ i}
$.
Then the value of $\delta_0  z$ is solution of the algebraic equation of the forth
power.
It can be easily solved.

When we take into
account only the leading term $f_{2\ j} z^4/4$ we come to the
method described earlier.

\subsection{Secondary nucleation at the increasing asymptotic}

The rate of nucleation
imaginary calculated without
droplets appeared through the second channel will be
called the ideal rate of nucleation.

All methods developed in two previous subsections can be used
here. But in the case of increasing
ideal rate of the droplets formation through the second channel
 there is a
following danger - the increasing density $n_i$ plays the role
analogous to the increasing ideal supersaturation in case of
dynamic conditions \cite{heterog}. So, at some density it is
possible to see the compensation of increase of density due to
evolution of droplets appeared through the first thermodynamic
channel by the molecules consumption  by the droplets appeared
through the second thermodynamic  channel. So, here we can see the
peak of intensive formation of droplets through the second channel
much earlier than we attain the asymptotic. We has to investigate this
situation.

The next difficulty is  connected with the fact that due to
increasing density the term $N_* (z +\delta)^3 $ can not be now
considered as the reason for  the
practically instantaneous  cut-off of the spectrum. This
radically completes the consideration.

We can introduce the natural length of spectrum $\Delta_{nat} z$
corresponding to the given vapor densities $n_{i\ fin}$. This length is
the length of spectrum when the initial vapor densities are
equal to $n_{i\ fin}$.
It is shown in \cite{book} that the characteristic time of the
spectrum formation in the situation of decay and in the situation of
dynamic conditions will be approximately one and the
same (see the law of inertia of characteristics of the phase
transition,
page 92). So, we have here some natural characteristic of duration
of the spectrum formation.

The characteristic time $\Delta_{nat} t$ will correspond to this
length of spectrum. One can compare $\Delta_{nat} t$ with the
relaxation length $\Delta_{rel} t \equiv \delta_{0\  n\ r} t  $ going from the
asymptotics.

When $\Delta_{nat} t >> \Delta_{rel} t$ one can easily use the
approach of the previous sections, there is nothing to change in the
previous approach.

When $\Delta_{nat} t << \Delta_{rel} t$ one has the situation
described as dynamic conditions with the source, which can be
effectively linearized (see \cite{book}). So, here the description
is also known (the only specific thing is that we have to add the
quantity $N_*$ to the number of droplets obtained at the second
peak of densities (the first peak is the initial peak)).

When $\Delta_{nat} t \sim \Delta_{rel} t$ one has use
 the modified Gaussian method, which has been already used to
investigate the nucleation on several types of heterogeneous
centers
in dynamic
conditions (see \cite{Gauss}).

The method similar to the
mentioned method
will be described below.
We shall present the simplest version which requires
the minimum formulas and calculations (but isn't too precise).
 We shall  describe here
the nucleation in dynamic condition with essentially non-linear
behavior of external supersaturation (effective external conditions)
in time (for external
supersaturation see \cite{Sevdin}).

We shall start with the case when the ideal supersaturation can be
linearized. The integral equation, which has to be solved can be
written in the following form
$$
 g(z)  = a \int_{-\infty}^z
(z-x)^3 \exp(c(\Phi(x) - \Phi_*) -  b g(x)) \ \ .
$$
Here $a$,
$b$, $c$ - are some constants, $\Phi$ can be interpreted as
$n_i$, index $*$ marks the value in some characteristic moment
$t_*$, which is the moment of maximum of $c(\Phi(z) - b g)$ and
$z(t_*) = 0$. The values $z$ and $x$ have the same sense as in the
previous section\footnote{The value $\Phi_*$ can be
taken as the asymptotic (without droplets appeared through the
second channel) value (without any correspondence
to the time $t_*$ \cite{PhysrevE94} ).}.

When $\Phi(z) - \Phi_*$ can be linearized the equation after the
certain renormalization can be written
$$
g(z)  = a'
\int_{-\infty}^z (z-x)^3 \exp(c'x -  b' g(x)) \ \
$$
with new values of parameters $c', b'$.

The last equation can be solved by the following approximate
procedure. We know that the iterations defined according
to
$$
g_{(0)} = 0 \ \ ,
$$
$$
 g_{(i+1)}(z)  = a' \int_{-\infty}^z (z-x)^3
\exp(c'x -  b' g_{(i)}(x)) $$ converge rather fast.
Already the spectrum with $g$ as the first iteration ensures the
correct number of appeared
droplets\footnote{Here it is taken into account that
$d^2 \Phi / dt^2 < 0 $.} (a relative error
is less\footnote{One
has to see this error when $\Phi - \Phi_*$ can be linearized
and then to prove that for $d^2 \Phi / dt^2 < 0 $ the error
will be smaller.} than
$0.15$). Why it takes place? Because the droplets appeared at the
first moments of time (when $b g$ was really small) are the main
consumers of vapor during the nucleation period. This takes place
due to the big power $3$ in  the integral term. So, there exists a
boundary $z=-b_0$, which have the following properties
\begin{itemize}
\item (')
The droplets appeared before $z=-b_0$
govern the evolution during the nucleation period.
 \item ('')
 The
inequality $ b g(-b_0)  \ll 1 $ takes place.
\end{itemize}

It is clear that $b_0$ takes the value, which has the order of $
c'^{-1}$ (more accurate it is $(0.7 \div 0.8)c^{-1}$).

So, we can suggest the following procedure:
At first to calculate the value
of $\tilde{g}$
defined as
$$
 \tilde{g} \equiv a \int_{-\infty}^{-b_0} dx
(z-x)^3 \exp(c(\Phi(z) - \Phi_*) ) \ \ .
$$
Then to calculate the
number of droplets as
\begin{equation} \label{nump}
N \sim a \int_{-\infty}^{-b_0} dz
\exp(c(\Phi(z) - \Phi_*) ) + a \int_{-b_0}^{\infty}  dz
\exp(c(\Phi(z) - \Phi_*) -  b \tilde{g}(z))
\end{equation}

This method is rather precise, the ratio $N_{appr}/ N$ where
$N_{appr}$ is the number of droplets calculated in (\ref{nump})
 and $N$ is precise solution is drawn
in\footnote{Here the linear case is considered.} Fig. 3

\begin{figure}

\begin{picture}(450,250)
\put(60,10){\special{em: moveto}}
\put(290,10){\special{em: lineto}}
\put(290,240){\special{em: lineto}}
\put(60,240){\special{em: lineto}}
\put(60,10){\special{em: lineto}}
\put(65,15){\special{em: moveto}}
\put(285,15){\special{em: lineto}}
\put(285,235){\special{em: lineto}}
\put(65,235){\special{em: lineto}}
\put(65,15){\special{em: lineto}}
\put(100,50){\vector(1,0){150}}
\put(100,50){\vector(0,1){150}}
\put(150,50){\vector(0,1){3}}
\put(200,50){\vector(0,1){3}}
\put(100,100){\vector(1,0){3}}
\put(100,150){\vector(1,0){3}}
\put(90,40){$0$}
\put(148,40){$1$}
\put(198,40){$2$}
\put(90,98){$1$}
\put(90,148){$2$}
\put(108,208){$N_{appr}/N  $}
\put(255,55){$-b_0 $}
\put(102.50,112.31){\special{em: moveto}}
\put(102.50,112.31){\special{em: lineto}}
\put(105.00,111.00){\special{em: lineto}}
\put(107.50,109.80){\special{em: lineto}}
\put(110.00,108.72){\special{em: lineto}}
\put(112.50,107.75){\special{em: lineto}}
\put(115.00,106.87){\special{em: lineto}}
\put(117.50,106.10){\special{em: lineto}}
\put(120.00,105.41){\special{em: lineto}}
\put(122.50,104.82){\special{em: lineto}}
\put(125.00,104.30){\special{em: lineto}}
\put(127.50,103.87){\special{em: lineto}}
\put(130.00,103.51){\special{em: lineto}}
\put(132.50,103.22){\special{em: lineto}}
\put(135.00,103.01){\special{em: lineto}}
\put(137.50,102.87){\special{em: lineto}}
\put(140.00,102.79){\special{em: lineto}}
\put(142.50,102.78){\special{em: lineto}}
\put(145.00,102.84){\special{em: lineto}}
\put(147.50,102.95){\special{em: lineto}}
\put(150.00,103.14){\special{em: lineto}}
\put(152.50,103.38){\special{em: lineto}}
\put(155.00,103.69){\special{em: lineto}}
\put(157.50,104.07){\special{em: lineto}}
\put(160.00,104.51){\special{em: lineto}}
\put(162.50,105.02){\special{em: lineto}}
\put(165.00,105.60){\special{em: lineto}}
\put(167.50,106.24){\special{em: lineto}}
\put(170.00,106.97){\special{em: lineto}}
\put(172.50,107.76){\special{em: lineto}}
\put(175.00,108.64){\special{em: lineto}}
\put(177.50,109.60){\special{em: lineto}}
\put(102.50,100){.}
\put(105.00,100){.}
\put(107.50,100){.}
\put(110.00,100){.}
\put(112.50,100){.}
\put(115.00,100){.}
\put(117.50,100){.}
\put(120.00,100){.}
\put(122.50,100){.}
\put(125.00,100){.}
\put(127.50,100){.}
\put(130.00,100){.}
\put(132.50,100){.}
\put(135.00,100){.}
\put(137.50,100){.}
\put(140.00,100){.}
\put(142.50,100){.}
\put(145.00,100){.}
\put(147.50,100){.}
\put(150.00,100){.}
\put(152.50,100){.}
\put(155.00,100){.}
\put(157.50,100){.}
\put(160.00,100){.}
\put(162.50,100){.}
\put(165.00,100){.}
\put(167.50,100){.}
\put(170.00,100){.}
\put(172.50,100){.}
\put(175.00,100){.}
\put(177.50,100){.}
\end{picture}

Figure 3.
Precision of the method in the linear case

\end{figure}

Now we shall apply this method to the case under consideration.

The equation we have to study is
$$
g_{2\ 1} =  f_{2\ 1} \int_{-\infty}^{z}
(z-x)^3 \exp(-\nu_{0\ c\ 2\ fin } |C_{ 0}| \exp(-x/\delta_0 z))
\exp(-A_1 g_{2\ 1}) dx \ \ ,
$$
where $\delta_0 z = \delta_{0 \
n\ 2} t \Lambda_{1\ 2}$ and $f_{ 2\ 1}$ has the sense $J_{ 2\ fin} $
divided by the rate of growth $\Lambda_{1\ 2}$.
Here $C_0 \sim  C_{n\ 0}$.
So, here $\Phi- \Phi_* \sim -\nu_{0\ 2\ c\ fin} |C_0 |
\exp(-x/\delta_0 z)$ and, thus,
$$
d\Phi / dx > 0 \ , \ \ \
 d^2 \Phi / d x^2 < 0 \ \ . $$

One has to add the term
$\exp(-\sum_j \nu_{j\ c\ 2} \frac{\xi_{j\ k\ fin}}{\xi_{1\ k\ fin}}
N_* ( z_{*}  +x)^3)
$
due to the initial peak of droplets. Here $z_{*}$ is is the coordinate of
this monodisperse peak at $t=t_*$. Then
\begin{eqnarray}
g_{2\ 1} =  f_{2\ 1} \int_{-\infty}^{z}
(z-x)^3 \exp(-\nu_{0\ c\ 2\ fin } |C_{ 0}| \exp(-x/\delta_0 z))
\nonumber
\\
\nonumber
\exp(-\sum_j \nu_{j\ c\ 2} \frac{\xi_{j\ k\ fin}}{\xi_{1\ k\ fin}}
N_* ( z_{*}  +x)^3)
\exp(-A_1 g_{2\ 1}) dx \ \
\end{eqnarray}
and inequality
$d^2 \Phi / d z^2< 0 $ also takes place.

Now
$$
c(\Phi(x) - \Phi) =
-\nu_{0\ c\ 2\ fin} |C_0| \exp(-x/\delta_0 z) - \sum_j \nu_{j\
c\ 2 } \frac{\xi_{j\ k\ fin}}{\xi_{1\ k\  fin}}
N_* ( z_{*}  +x)^3
$$

Due to the  last inequality  the method will work even better
than in the linear case because here the properties (') and ('')
will be better. The problem is whether one can calculate
$\tilde{g}$ analytically.

Actually $\tilde{g}$ is the polynomial in $z$ of the third power
$$ \tilde{g} = \sum_{i=0}^3 z^i (-1)^{3-i}w_i \frac{3!}{i!(3-i)!}
\ \ \ ,  $$ $$ w_i =a \int_{-\infty}^{-b_0} x^{(3-i)}
\exp(c(\Phi(z) - \Phi_*) ) \ \ .  $$

Now we shall
calculate $N$ and have to take definite integrals. It can be done
by the quickest descent method since in the subintegral expressions
there are only exponents.

\section{Discussion}

The theory for multicomponent condensation can not be presented
in such a concrete and compact form as in the case of one component
condensation. This is a result of the presence of complex and
generally arbitrary coefficients of activity $f_i$. Here we have
supposed only some general properties of $f_i$ such as the
regularity of solution at small concentrations

The condensation in the binary systems is much more interesting
than in the one component case. The properties  of reverse
condensation, of the secondary nucleation are simply absent in the
one component case. Now they are predicted only theoretically  and
it will be very interesting  to see them experimentally.
Unfortunately, even the
experimental results used in \cite{djikaiev2} don't belong to the
kinetics of the whole phase transition, but to the  stationary
rate of nucleation. The last is certainly the important
characteristic
but it has no direct connection
with the specific phenomena of the multicomponent kinetics,
which is the central point
discussed here.  This publication has to
stimulate the experimental investigations in this direction
because now it is absolutely clear  what effects and
characteristics are worth searching.

In the last section the description of the "long nucleation" was
done only for the period of  nucleation and for the two
component case. As for the  description of the second period it
can be done in the same manner as for the case of the
ordinary
nucleation. It is necessary only to consider the two channels of
nucleation in all places  where  we speak about the initial stage
of condensation and about the asymptotic stage of condensation.
The situation when the evolution of the droplets from the first
channel already attains the asymptotic stage (that's why we prefer
to speak about stages of condensation instead of the periods of
condensation) and the droplets\footnote{We have to
consider the initial, intermediate and the final periods
with reference for every component.} from the second channel are still
at the initial stage is rather specific. But
still in frames of approximations of initial stage and final stage
it is easy to give the adequate description.

For different  components the  power of consumption
 of the surplus vapor can be different. That's why at the
given moment of time
the different components can be at different stages. The evolution
becomes unclear. In this context it is important
to see the property of the global avalanche consumption, which
states that the
intermediate period has relatively (in comparison with
the duration of the first period) short duration. Then
for every concrete complex
system one can act in frames of simple solutions and get the simple
true kinetics of nucleation.

Here we shall briefly summarize the new items proposed in this
paper:
\begin{itemize}
\item
The consideration of kinetic and
thermodynamic channels is given.
\item
The correct description of
the relaxation to the stationary state in the channel
  is given.
 \item
Essential  errors of the elementary
description of nucleation from \cite{djikaiev2} are corrected.
\item
The description of the condensation of the main quantity of
substance is given. This description corrects the description
proposed in \cite{djikaiev2}. It is shown that the precise
solution is absent. Several
effective approximations are suggested. They are
rather effective.
\item
The situation of the "long nucleation" is
analyzed. The methods to give the adequate description of this
situation are proposed.

\end{itemize}

But the main result of the article is the complete analytical
description of the condensation process in the system with a
mixture of substances in a metastable phase. All main
characteristics of the process such as the numbers of droplets of
dufferent sorts and the times of duration of nucleation in different
channels of nucleation are determined explicitly.

We have not given the estimates for the beginning of the
transformation of the condensation process into the process of
coalescense (super-condensation). The necessary estimate
comes from condition  that
the coordinate of monodisperse spectrum $\nu_i$ has to be $2$ or
$3$ times greater than the critical size $\nu_{i\ c}$ in the every
corresponding channel.  The announced estimate
can be easily checked on the base of obtained solution, which gives
the estimate for the time of the end of the condensation
process\footnote{ Also the statement from
\cite{djikaiev2} that the theory of multicomponent coalescence is
absent is wrong - the results presented in \cite{kukushkin} cover
the case of multicomponent coalescence.}.

The generalization of the description of the "long nucleation" for
the multicomponent case is rather simple. It is clear that in the
$n$-component mixture condensation the maximum number of separate
nucleation processes is limited by $n$. So, we have to act in the
following manner: Consider the first channel of nucleation, then
see what components aren't exhausted (the corresponding
concentrations in kinetic channels are small). If there are some
unexhausted components we can search for the channels where only
these unexhausted components are the leading ones (the
concentrations of all other components at thermodynamic channel
are small). We shall do it in the sequence of increasing heights
of activation barriers of nucleation. May be we shall find the
channel with a necessary properties. This channel may not be  the
second in the total sequence  of the activation barriers heights,
but all channels before this channel are now out of action.
 The nucleation through this channel will
be the secondary nucleation and it is already described. Then we
shall reconsider the set of unexhausted components (these
components are the components, which are small in both kinetic
channels) and seek for the kinetic channel with essential
concentrations (concentrations, which aren't small) from the set of
previously unexhausted components. This channel will be the
leading
one. This completes the loop, which can be repeated again. The
number of such loops is limited from above by $l$.

As the result we have the set of channels, which will be called as
essential channels (both thermodynamic and kinetic). The
nucleation in  all channels from this set can be described
analogously to the two component case. There is only one
difficulty, which is specific to the multi component case. In the two
component case the consumption of vapor by the droplets from the
first channel occurs in the pseudo one-dimension manner (the
concentration of one component is small, so the concentration of
the other component is known also and, thus, all concentrations
are known; moreover the value of the droplets volume is determined
only by one component). In the multicomponent case there will be a
real problem to describe this evolution.  But with the help of
methods presented in the section devoted to the second period this
situation can be described analytically also.

To give the necessary description one can split the evolution into
the initial stage and the asymptotic stage.  Since for droplets
appeared from one channel the evolution is already at the final
asymptotic stage and for the droplets from the other channel the
evolution is at the initial stage at one and the same moment then
it is preferable to speak about stages instead of periods.   The
initial stage and the final stage practically cover all evolution.
This is the crucial point of description. Then it is rather easy
to take into  account the influence of the already existing spectrums
of droplets appeared through the already closed channels, because
we know the functional expressions for them as  functions of
time\footnote{It can be necessary to add some terms, which are similar to
$N_* t^3$ (the coefficient will be another) and some terms at the
asymptotic (here the structure is also known).}. So, the functional
form of equations will be the same as for the binary system and it
is easy to get the approximate solution
in  analytical form.

One have
also to keep in mind that the final state corresponding to the
different channels can be different. Certainly, it doesn't produce
essential
difficulties.

\end{document}